\documentclass{article}

\usepackage{mathalpha}
\usepackage{amssymb}
\usepackage{fullpage}
\usepackage{amsmath}
\usepackage{booktabs}
\usepackage{enumitem}
\usepackage[normalem]{ulem}
\usepackage{graphicx}
\usepackage{algorithm,algorithmicx,algpseudocode}

\usepackage{multirow}
\usepackage[notquote,notcomma]{hanging}
\usepackage{bm}

\usepackage[sort&compress,comma]{natbib}
\usepackage[colorlinks=true, allcolors=violet, backref = page]{hyperref}
\renewcommand*{\backref}[1]{}
\renewcommand*{\backrefalt}[4]{
    \ifcase #1 (Not cited.)
    \or        (Cited on page~#2.) 
    \else      (Cited on pages~#2.) 
    \fi}

\usepackage{endnotes}
\usepackage{float}
\floatstyle{plaintop}
\restylefloat{table}
\restylefloat{figure}
\usepackage{caption}

\usepackage{color, colortbl}
\usepackage{xcolor}
\definecolor{Gray}{gray}{0.9}

\usepackage{amsmath}    
\usepackage{amssymb,mathrsfs,xargs}
\newcommand{\indep}{\mathcal{I}}

\newcommand{\sIndep}{S_\indep}
\newcommand{\sStar}{S_\star}
\newcommand{\graph}{\mathbb{G}}

\newcommand{\minSecG}{\mathcal{G}}

\newcommandx{\securityState}{\Upsilon}
\newcommandx{\security}[2][1=S,2=i]{\securityState_{#1}^{#2}}
\newcommand{\coreSubset}{\mathbf{\mathcal S}}

\renewcommand{\hat}{\widehat}
\renewcommand{\bar}{\overline}
\renewcommand{\tilde}{\widetilde}

\usepackage{graphicx}   
\usepackage{verbatim}   
\usepackage{color}      
\usepackage{hyperref}   
\usepackage{epsfig}
\usepackage{eqnarray,amsmath}
\usepackage{eqnarray,amssymb}
\usepackage{mathtools}
\usepackage{enumitem}
\usepackage{setspace}

\usepackage{empheq}
\usepackage[english]{babel}

\usepackage{float}
\usepackage{breqn}
\usepackage{graphicx}
\usepackage{tabularx}
\graphicspath{ {images/} }


\everymath{\displaystyle}

\newtheorem{theorem}{Theorem}[section]
\newtheorem{lemma}[theorem]{Lemma}
\newtheorem{example}[theorem]{Example}
\newtheorem{proposition}[theorem]{Proposition}
\newtheorem{corollary}[theorem]{Corollary}

\newcommand{\Halmos}{\hfill$\blacksquare$}

\usepackage[english]{babel}
\usepackage[nameinlink]{cleveref}

\crefname{figure}{Figure}{Figures}
\crefname{lemma}{Lemma}{Lemmata}
\crefname{proposition}{Proposition}{Propositions}
\crefname{theorem}{Theorem}{Theorems}

\usepackage{enumitem}

\renewcommand\footnoterule{{\color{black}\hrule height 0pt}}

\usepackage{hyperref}

\usepackage{graphicx}
\graphicspath{ {images/} }

\usepackage{natbib}
 \bibpunct[, ]{(}{)}{,}{a}{}{,}

\DeclareMathAlphabet\mathbfcal{OMS}{cmsy}{b}{n}

\makeatletter

\colorlet{rosso}{red!90!white}
\colorlet{bloo}{blue!90!white}

\definecolor{mycolor}{RGB}{200,0,0}

\renewcommand{\footnoterule}{
  \kern -3pt
  \hrule width \textwidth height 0.25pt
  \kern 2pt
}

\title{Cooperative Security Against Interdependent Risks}
\author{Sanjith Gopalakrishnan$^\star$ \and Sriram Sankaranarayanan$^\dagger$}
\date{\small
	$^\star$Desautels Faculty of Management, McGill University\\
	$^\dagger$Indian Institute of Management Ahmedabad
}

\begin{document}

\maketitle

\begin{abstract}
Firms in inter-organizational networks such as supply chains or strategic alliances are exposed to interdependent risks. These are risks that are transferable across partner firms, such as contamination in food supply chains or data breaches in technology networks. They can be decomposed into intrinsic risks a firm faces from its own operations and extrinsic risks transferred from its partners. Firms broadly have access to two security strategies: either they can independently eliminate both intrinsic and extrinsic risks by securing their links with partners, or alternatively, firms can cooperate with partners to eliminate sources of intrinsic risk in the network. We develop a graph-theoretic model of interdependent security and demonstrate that the network-optimal security strategy can be computed in polynomial time. Then, we use cooperative game-theoretic tools to examine whether and when firms can sustain the network-optimal security strategy via cost-sharing mechanisms that are stable, fair, computable, and implementable via a series of bilateral cost-sharing arrangements. We consider different informational assumptions in the network and show that, in the private information setting where players know only their own costs, firms have a clear incentive to cooperate globally whereas, in the presence of public information, there may not exist cost-sharing mechanisms that can sustain network-wide cooperation. We then design a novel cost-sharing mechanism: a restricted variant of the well-known Shapley value, the {\it agreeable allocation}, that is easy to compute, bilaterally implementable, ensures stability, and is fair in a well-defined sense. However, the agreeable allocation need not always exist. Interestingly, we find that in networks with homogeneous cost parameters, the presence of locally dense clusters of connected firms precludes the existence of the agreeable allocation, while the absence of sufficiently dense clusters (formally, $k$-cores) guarantees its existence. Finally, using the SDC Platinum database, we consider all inter-firm alliances formed in the food manufacturing sector from 2006 to 2020. Then, using simulated cost parameters, we examine the practical feasibility of identifying bilaterally implementable security cost sharing arrangements in these real-world alliances.
\end{abstract}

\section{Introduction and Related Literature} 
\medskip 

Firms increasingly belong to a variety of inter-organizational networks, such as complex supply chains, strategic alliances, or other types of partnerships. Membership in these networks can evidently yield economic benefits, but they also necessitate substantial additional security investments due to increased exposure to interdependent or contagion risks \citep{kunreuther2003}. For instance, in January 2013, the European food industry endured a horse-meat contamination scandal \citep{lawrence2013}. Meat products from several retailers and fast-food chains in the United Kingdom and Ireland, advertised as containing beef, were discovered upon testing to have been contaminated with horse-meat. Further investigation revealed that in the complex meat supply networks, with contractors and subcontractors spread all across Europe, a particular supplier had indulged in deliberate contamination in a bid to cut costs. Several retailers, including Britain's largest retailer, TESCO, that had sourced the contaminated meat, faced economic repercussions from a drop in sales and reputational harm. Other notable cases of supply contamination include the adulteration of milk with melamine \citep{mu2016, levi2020} and the 2008 heparin adulteration scandal \citep{babich2012}. Contamination in supply networks, upon discovery, typically results in product recalls, regulatory fines, and brand equity loss, often entailing substantial costs for the concerned firms. 

Besides supply networks, interdependent risks can arise in other contexts too. For instance, businesses have a growing recognition that they bear a social responsibility to secure their consumer data from cyber threats \citep{pollach2011}. Malware infecting the systems of a company in an inter-firm network can gain access to the IT systems of its partner firms. Due to poor cyber-security practices by partner firms, companies such as Target and Home Depot have been the victims of high-profile data and privacy breaches \citep{mcafee2015}. In today's highly interconnected networks, risks like contamination in food supply chains or consumer data breaches assume an interdependent nature. That is, the risks faced by a firm depend not only on internal risks arising from their own operations but also on the risk transferred from partner firms in the network. Further, the above examples involve risks transferred between networked partners with ongoing and frequent repeated interactions. Thus, a firm vulnerable to internal risks is near-certain to transfer this risk to its partner firms if these partners do not take appropriate remedial actions.

Therefore, to secure themselves against interdependent risks, two general strategies are available to networked firms. First, firms in the network can choose to invest cooperatively in securing themselves, thereby removing sources of risk. Second, alternatively, firms can choose to independently secure themselves by eliminating risk from internal operations and then investing in security across the links that connect them to the other firms in the network. So, for example, firms could cooperatively share the costs of supplier quality improvements, thereby investing in suppliers' embracing responsible operational practices. Alternatively, a retailer can implement quality standards for internal processes, and simultaneously, inspect and quality test incoming products supplied by direct partners. The latter would correspond to the {\it independent security strategy}, while the former corresponds to the {\it cooperative security strategy}. 

Security against interdependent risks is associated with positive externalities since other firms are benefited from the presence of a secured firm in the network. This would intuitively suggest that cooperative network-wide security against interdependent risks can be a cost-effective strategy as compared to each firm in the network independently securing itself. However, cooperation can be hindered by disagreements over cost-sharing arrangements. Firms, in general, are heterogeneous, both, in the costs they incur to secure themselves as well as in the penalties that they may face in case of a realized risk. Thus, a priori, it is not clear whether there will always exist a stable and fair sharing of security costs that can sustain network-wide cooperation. Furthermore, networked firms typically have visibility and mechanisms to cooperate and monitor with only immediate partners. For instance, extended multi-tier supply chains are often associated with a loss in visibility over firms further away in the network \citep{caro2021}. Thus, it is also unclear whether one can find suitable mechanisms to implement cost-sharing arrangements that circumvent coordination across firms that are not immediate or direct  partners.\footnote{Relatedly, \citet{dawande2021}, in a review of recent research on socially responsible operations management, note that, {\it "a topic that has not received much attention yet is the design of cooperative strategies among stakeholders in different tiers of a supply chain to collectively ensure socially responsible actions across the supply chain... the utilities of different players from actions such as auditing, inspections, and testing become interconnected in a complex manner. Consequently, the sharing of costs in a fair manner to incentivize cooperation across tiers becomes challenging."}} 

To address these issues, in this paper, we consider an interdependent security model on a network and an associated cost-sharing game. In our model, as motivated above, firms face an intrinsic risk from their internal operations and an extrinsic risk from their unsecured partners in the network. Firms in the network are heterogeneous in the costs they incur to secure themselves and the penalties they face in case of an actualized threat. 

Further, we also consider our network security model under differing informational assumptions. In our {\it private information model}, we assume that all cost parameters are privately known to players. So, in the absence of explicit cooperation, each firm's security actions cannot be observed or inferred by other firms in the network. This private information assumption is a marked distinction from existing models of interdependent security in the literature, which typically assume that various model parameters and actions are public information. In several real-world contexts, in the absence of formal mechanisms for cooperation, firms are neither aware of the security efforts undertaken by other firms nor can they infer their efforts since the underlying cost structures are typically private information. However, in certain other scenarios, it would be more reasonable to assume that firms are indeed aware of the security costs of other firms in the network. Therefore, we also analyze our network security model with the alternative informational assumption wherein efforts and cost structures are {\it public information}. Further, studying these two extreme informational assumptions also permits us to separate the benefits of cooperation arising from interdependence and information acquisition. In the e-companion, we also consider a more general hybrid model, the {\it partial information model}, where, as in practice, due to regulatory requirements or strategic disclosures, the cost parameters and efforts of some firms are publicly known whereas the costs and efforts of other firms are only known privately. 

The network-optimal security strategy under all informational assumptions is identical, and we demonstrate that it can be computed in polynomial time using a minimum weighted cut network-flow algorithm. Then, we adopt a cooperative game-theoretic approach to assess whether agents have an incentive to cooperate across the entire network and share the security investment costs. We show that, under the private information setting, agents have a clear incentive to cooperate globally, i.e., form the grand coalition and share the resulting security costs. However, with even some information being public in the network, we show that, in general, there do not exist cost-sharing mechanisms that can ensure the stability of the grand coalition. This can be explained by two drivers: first, with public information, the benefit from additional
information acquisition is lowered. Thus, the benefits from cooperative security in the public
information setting are arguably lower. Second, public information engenders free-riding since
firms can now anticipate and observe the security actions of other firms in the network and
benefit from the cooperation of other firms in the network without participating in the grand
coalition and sharing security costs. In similar cooperative settings with externalities, free-rider concerns are acknowledged as a fundamental reason often precluding the stability of the grand coalition (see, e.g., \citet{yi1997stable}).

Importantly, we then introduce the notion of {\it bilateral implementability}. A cost-sharing arrangement is said to be bilaterally implementable if it can be enforced by a series of bilateral cost-sharing agreements between only direct partners in the network. Bilaterally implementable cost-sharing mechanisms are resistant to the aforementioned limitations of network visibility and control. It is generally assumed, for example, in managing supply chains that it is easier for firms to contract with their immediate suppliers with whom they share direct relationships and that it is more challenging to gain visibility, manage, and contract with deep-tier suppliers (see, e.g., \cite{huang2020} and \cite{dong2022blockchain}). We propose a novel security cost sharing mechanism, the {\it agreeable allocation}, which is a restricted variant of the Shapley value allocation \citep{shapley1971}. We then demonstrate that the agreeable allocation satisfies notions of stability, is formalizably fair, and unlike the Shapley value, is easily computable, and always bilaterally implementable. However, the agreeable allocation may not always exist. We then construct $\delta$-agreeable allocations that satisfy a generalized notion of ($\delta$+1)-lateral implementability, for an integer $\delta \geq 1$, whereby firms that are at a distance of at most $\delta$ from each other in the network can enter into cost-sharing agreements. When $\delta = 1$, we recover bilateral implementability. This allows us to delineate a hierarchy of cost-sharing mechanisms such that as $\delta$ increases (i.e., firms that are farther away from each other in the network are allowed to cooperate), the corresponding $\delta$-agreeable allocation is more likely to exist. 

To analyze the effects of network structure on the existence of the agreeable allocation, we consider the special case of quasi-homogeneous networks, i.e., networks where the security cost parameters are equal. We then provide a structural graph-theoretic characterization for the existence of the agreeable allocation in these networks. Specifically, we show that the local density of networks plays a key role in determining whether the agreeable allocation exists.

In summary, one can view our work in both descriptive and normative terms. Descriptively, we observe that network-wide security cooperation is efficient and in some cases, this cooperation can be sustained with suitable cost-sharing arrangements. However, when concerns pertaining to computability and implementability of these cost-sharing mechanisms are incorporated, network-wide security cooperation is rendered more challenging. Normatively, via our analysis of the agreeable allocation and its extensions, we are able to provide insights into when and how these implementation challenges can be surmounted. 

\medskip 
\subsection{Overview of Related Literature}
\medskip 

This work is related to three distinct streams of literature. First, it contributes to extant work on social responsibility and risk management in supply chains. Second, our work is closely tied to interdependent security models introduced by \cite{kunreuther2003}. One of our aims is to bridge these two bodies of literature. Finally, our work adds to the growing literature on applications of cooperative game theory to operations management.  
\smallskip 
\subsubsection*{Supply Chain Social Responsibility and Risk Management.}

There is a vast literature investigating the role of several instruments such as auditing \citep{plambeck2016, caro2018, fang2020, chen2020}, inspection and testing \citep{babich2012, lee2018}, and more recently, contracts \citep{dhingra2021},  in mitigating social responsibility risks associated with extended global supply chains. We refer the interested reader to \citet{dawande2021} for a recent review. While previously, most of this literature dealt with two firm or dyadic scenarios, recently, several studies also deal with multi-tier supply chains, e.g., supply networks with three tiers or other network structures \citep{huang2020, zhang2021, chen2020}. Also closely related to our work, \citet{feng2021} study the implementation of ESR programs in general supply networks and gain sharing via a bilateral bargaining framework that generalizes a conventional Shapley value based cooperative-game theoretic approach. Recently, \citet{blaettchen2021traceability} also study the optimal adoption seeding of traceability technologies which carry several implications for sustainable practices in supply networks. While we view our work as contributing to this stream of literature, we note that it bears some differences. For instance, we consider a general network structure and do not impose any structural assumptions. Second, our work deals with only interdependent risks. That is risks that are contagion risks spreading via the network. These scenarios include cases such as food contamination risks or data breach threats as motivated in the introduction. 
\smallskip 

\subsubsection*{Interdependent Security.} In terms of model development, our work is most closely related to the interdependent security literature. Interdependent security models were introduced by \citet{kunreuther2003} and have since spawned a rich literature in the intersection of economics and computer science that studies various related models (see, for example, \citet{laszka2014} for a review). In these models, as in ours, the security of agents depends on an agent's own actions (direct risk, or as we term it, intrinsic risk) and those of other agents (indirect or extrinsic risk). The present work aims to bridge the interdependent security literature with the rich stream of work on socially responsible operations in supply networks. While this research stream inspires our model, our work differs from existing literature in some crucial ways. First, in several of the existing models, the agents can only curb their own intrinsic risk and cannot mitigate extrinsic risks. Second, a majority of the interdependent security literature adopts a non-cooperative (game-theoretic) perspective. They assume that players in the network act to secure themselves independently and then characterize and compute the non-cooperative equilibria of these games. \citet{kearns2003} and \citet{chan2012} develop algorithms to compute the equilibria of classes of interdependent security games. \citet{heal2007} also consider the Nash equilibria of such games and study conditions to tipping sub-optimal equilibria to an optimal one. \citet{chan2014} consider a more general model where agents can influence the transfer of extrinsic risk and then analyze equilibria computations. However, this literature largely ignores issues of cooperation in networks and the problem of when and how cooperation can be sustained. In practice, agents can and indeed do cooperatively secure themselves against interdependent risks. This, therefore, is the central focus of this present paper.  

\subsubsection*{Cooperative Game Theory in Operations Management.} Finally, we also contribute to the growing body of work dealing with applying cooperative game theory to problems in operations management.  For a review of this literature, we refer the reader to \citet{nagarajan2008}. Benefits of cooperation can be realized and therefore studied in several diverse settings. Some recent applications include inventory pooling \citep{kemahliouglu2011}, inventory transshipments \citep{granot2003, sosic2006}, demand information sharing \citep{leng2009}, supplier alliances to mitigate order default risk \citep{huang2016}, production schedule coordination \citep{aydinliyim2010}, supply chain emissions management and reduction \citep{gopalakrishnan2021a, gopalakrishnan2021b}, recycling \citep{gui2018, tian2020}, humanitarian operations \citep{ergun2014}, vaccine distribution \citep{westerink2020} and so forth. Related to our work, \citet{mu2019} study quality management in milk cooperatives. In dairy cooperatives, individual farmers can shirk on quality and free-ride on the higher quality milk produced by other farmers in the cooperative. \citet{mu2019}, therefore, develop a revenue allocation rule that achieves quantity and quality efficiency with minimal testing while incorporating other practical implementation considerations. 

\section{A Network Security Model} 
\medskip 

We consider a set of heterogeneous players\footnote{The terms {\it agents}, {\it firms}, and {\it players} are used interchangeably in this paper.} denoted by $N$. 
Following standard graph-theoretic notation, let us suppose that the players occupy a network denoted as $\graph = (N, A)$. 
The node set $N$ of the network coincides with the set of players with each player occupying a unique corresponding node in $\graph$. 
An arc $(i, j) \in A$ for $i, j \in N$ represents a directed link from the player $i$ to the player $j$. 
The set of arcs in the network is denoted by $A$. 
Let $N^+(i)$ denote the set of players in $N$ to which $i$ is connected by an outgoing arc $(i,j) \in A$, and similarly, let $N^-(i)$  be the set of players $j \in N$ such that the arc $(j,i) \in A$. 
Further, let $N(i) := N^+(i) \cup N^-(i)$. 

Each player faces two independent sources of risk: an {\it intrinsic risk} from its own operations and an {\it extrinsic risk} transferred from its partnerships with unsecured players.\footnote{In the interdependent security literature, intrinsic and extrinsic risks are sometimes referred to as {\it direct} and {\it indirect} risks, respectively.} We assume the cost incurred by player $i$ to secure itself against intrinsic risks is given by $\theta_i$. 
Further, the cost incurred by $i$ to secure itself against the extrinsic risk transferred  from a partner in the network $j$ is denoted by $\xi_{ji}$.  
Each player $i$ exerts binary actions, $x_i \in \lbrace 0, 1\rbrace$, and $y_{ji} \in \lbrace 0,1\rbrace$ for all $j \in N^-(i)$, corresponding to whether to secure itself against its own intrinsic risk and extrinsic risk from its partners, respectively. 
Since different players may face differing penalties (in regulatory fines or reputational damage) in the case of a realized risk, we assume an {\it unsecured} player $i$ faces an expected penalty of $L_i$. 
A {\it secured} player faces a zero penalty. We will subsequently clarify when a player is said to be {\it secured} and {\it unsecured}, respectively.   

As outlined in \S1, firms can derive two distinct advantages from cooperative security in networks: first, the benefit of interdependence, which involves internalizing the positive externality of security, and second, the advantage of information acquisition. Accordingly, we first consider two extreme informational assumptions, a {\it private information} model where each player, in the absence of cooperation, is aware of and can observe only its own security cost parameters and actions. At the other extreme, we also consider the more traditional informational assumption of {\it public
information} where, even in the absence of cooperation, each player can observe the costs and actions of all other players in the network. 

\subsubsection*{Private Information Model.} In the private information model, we assume that all cost parameters including the cost of securing against intrinsic risk, $\theta_i$, and the expected penalty in case of a realized risk, $L_i$, are private information known only to player $i$. 
Similarly, the cost, $\xi_{ji}$, to secure the directed link between players $j$ and $i$ is assumed to be known only to players $i$ and $j$. 
This private information assumption is a departure from several existing models of interdependent security.  Specifically, the private information assumption implies that in the absence of explicit cooperation between players $i$ and $j$, neither can observe or infer the actions of the other. Thus, in this scenario, we can formally define the {\it information set} of a player $i$ acting independently as $I(i, \{i\}) = \lbrace \theta_i, \xi_{ij}, \xi_{ji}, L_i, x_i, y_{ji} : j \in N^-(i)\rbrace$.
Therefore, in this scenario, the information set of player $i \in N$ who cooperates with the set of players $i \in S \subseteq N$ expands and is given by $I(i,S) = \cup_{j \in S}\  I(j,\{j\}) = \lbrace \theta_j, \xi_{kj}, \xi_{jk}, L_j, x_j, y_{kj} : j \in S, k \in N^-(j)\rbrace$.

\subsubsection*{Public Information Model.} In contrast, in the public information model, we assume that all firms can observe each other's cost parameters and security actions even in the absence of cooperation. Then, the information set of a player $i$ acting independently is $I(i, \{i\}) = \lbrace \theta_j, \xi_{jk}, \xi_{kj}, L_j, x_j, y_{jk} : j \in N, k \in N^-(j)\rbrace$. Therefore, in the public information scenario, $I(i, S) = I(i, \{i\})$, and  firms upon cooperation do not derive any benefits from additional information acquisition. By analyzing and comparing these two extreme informational assumptions, we can comment on the benefits from cooperation along the two dimensions of interdependence and information acquisition. 

\subsubsection*{Partial Information Model.} In practice, even in the absence of explicit cooperation, the security costs and actions of certain firms may be public knowledge, due to regulatory requirements or strategic disclosures, whereas the costs and actions of other firms may only be known privately. Thus, we also consider a more general {partial information model} which assumes that the costs and actions of a subset of firms, $\mathcal{P} \subseteq N$ are publicly known to all firms in the network whereas the costs and actions of firms in $N\backslash \mathcal{P}$ are only privately known. Therefore, in this scenario, $I(i, \{i\}) = \lbrace \theta_j, \xi_{jk}, \xi_{kj}, L_j, x_j, y_{jk} : j \in \mathcal{P}\cup\{i\}, k \in N^-(j)\rbrace$. This more general hybrid model subsumes both the private and public information models described above. Clearly, when $\mathcal{P} = \emptyset$ and $\mathcal{P} = N$, we recover the private and public information models, respectively. In the interest of expositional clarity and brevity, we consider the private information and public information models in the paper and extend the discussion to the general partial information model in the e-companion \S EC.4. 

\subsubsection*{Security Actions.} Players in the network choose security actions, $x_i \in \lbrace 0, 1\rbrace$, and $y_{ji} \in \lbrace 0,1\rbrace$ for all $i \in N$ and $j \in N^-(i)$ after considering the relevant trade-off between the costs of security and the expected penalty in case of a realized risk. In order to do so, each player first forms beliefs on the {\it security states} of other firms in the network. That is, a player $i$, cooperating with players in $S$ and with the information set $I(i,S)$, forms a belief on the security state of $j \in N$ denoted by $\sigma_{ji}(I(i,S)) \in \lbrace 0, 1\rbrace$ where $\sigma_{ji} = 0$ means player $i$ believes $j$ to be unsecured, and if  $\sigma_{ji} = 1$, then $i$ believes $j$ is secured. We will subsequently clarify how players form beliefs on the security states of other firms in the network. Then, player $i$ chooses security actions $x_i$ and $y_{ji}$ accordingly to determine its own security state based on its beliefs. Since interdependent risks are transferable across partners, a player $i$ identifies itself as secured, i.e., $\sigma_i = 1$, if and only if its secured against its own intrinsic risk, i.e., $x_i = 1$, and further, is also secured against extrinsic risks, i.e., $y_{ji} = 1$ for all players $j \in N^-(i)$ who it believes to be unsecured. For clarity, we note that the security state $\sigma_i$ of player $i$ as a function of its own security actions, given its information set and its beliefs on the security states of its network partners, satisfies the following,

\begin{equation} 
\sigma_i(x_i, {\boldsymbol{y}_i} \vert I(i,S)) =  
\begin{cases}
    0, &\text{ if } x_i\prod_{\substack{j \in N^-(i)\\ \sigma_{ji} = 0}}y_{ji} = 0,\\ \\
    1, &\text{ otherwise.}\\
  \end{cases}
  \label{eqn2}
\end{equation}

Thus, the expected security cost incurred by a player $i$ is given as follows,
 \begin{equation}\label{eq:Cost}
  U_i(x_i, {\boldsymbol{y}_i}| I(i,S)) =  L_i(1-\sigma_i(x_i, {\boldsymbol{y}_i}| I(i,S)))+ \theta_ix_i + \sum\limits_{{j \in N^-(i)}} \xi_{ji}y_{ji}.
 \end{equation}

The first term in (\ref{eq:Cost}) corresponds to the expected penalty from a realized risk and is incurred only when the player $i$ is unsecured. The second and third terms correspond to the costs of securing itself against intrinsic risks, and extrinsic risks from unsecured partners, respectively.  

In \S3 and \S4, we analyze cooperative security strategies and the associated security cost sharing problem in the private information model whereas in \S6, we study the public information model. This sequence is chosen for expositional clarity. Further, in the interest of parsimony, we relegate the analysis under the general {\it partial information} model where each player acting independently is aware of the cost parameters and actions for only a subset of players to the e-companion \S EC.4.

\section{Security Strategies under Private Information}
\medskip 

Under the private information assumption, since a player cannot observe or infer the security actions of other players, we assume a player $i$ forms a worst-case belief on the security states of players it does not explicitly cooperate with.  That is, a player $i$ cooperating with the set of players $S \subseteq N$ forms the worst-case belief that $\sigma_{ji} = 0$ for all players $j \notin S$. Therefore, $i$ identifies itself as secured if and only if it is secured against its own intrinsic risk, $x_i = 1$, and further, is also secured against extrinsic risks, $y_{ji} = 1$ for all $j$ such that $\sigma_{ji} = 0$, i.e., (i) for $j$ not in $S$, and (ii) for $j$ in $S$ who are themselves not secured. Therefore, in the private information model, the security state of $i$ is denoted by $\sigma_i \in \lbrace 0, 1\rbrace$, where $\sigma_i = 0$ means, in the worst-case, player $i$ is unsecured, and if $\sigma_i = 1$, then $i$ is secured in the worst-case. Similar worst-case considerations are commonly employed in diverse network security applications  (see, e.g., a review on planning for supply network disruptions by \cite{snyder2006planning}).

We now consider two forms of security strategies in the network: the {\it independent security strategy} and the {\it network-optimal security strategy}. While the former corresponds to the no-cooperation, i.e., individually rational scenario, the latter corresponds to the full-cooperation, i.e., the network-optimal situation. In \S4, we will consider all intermediate cooperative security strategies, i.e., where a subset of firms in the network cooperatively secure themselves. 
\medskip

\subsubsection*{Independent Security Strategy.} Since the players are not cooperating with each other on their security actions, as noted previously, the information set of each player $i \in N$, $I(i,\{i\})$, only contains its own actions, expected penalty, and security costs. Then, player $i$ is said to be independently secured if $U_i$, as defined in (\ref{eq:Cost}), is minimized when $\sigma_i = 1$, for a suitable choice of $x_i$ and ${\boldsymbol{y}_i}$. The set of all players in $N$ which are independently secured is denoted by $\sIndep$. The following proposition characterizes when a player is independently secured. All proofs are provided in the e-companion.

\begin{proposition}\label{prop_independent}
A player $i \in \sIndep$ if and only if $L_i \geq \theta_i + \sum\limits_{j \in N^-(i)} \xi_{ji}$. Further, then, $x_i = y_{ji} = 1$ for all $j \in N^-(i)$.
\end{proposition}

The above proposition captures two straightforward notions in the private information setting: (i) the independent security strategy is based on a simple trade-off between the cost of security and the expected penalty incurred from not securing itself, (ii) for an agent acting independently, it is not optimal to partially invest in securing some links and not others.

\subsubsection*{Network-Optimal Security Strategy.}  In this setting of full network-wide cooperation, the information set of each player contains all the security costs and expected penalties of all other players in the network. The players act to minimize the total expected security cost of the network.
\begin{align}\label{eq:GraphCost}
U(\graph) &=\ \min_{\boldsymbol{x},\boldsymbol{y}}\quad  \sum_{i \in N} U_i(x_i, {\boldsymbol{y}_i}| I(i,N))   \nonumber \\
 &=\  \min_{\boldsymbol{x},\boldsymbol{y}}\quad   
 \sum_{i \in N} 
 \left( 
    L_i(1-\sigma_i(x_i, {\boldsymbol{y}_i}| I(i,N))) 
    + \theta_ix_i 
    + \sum_{j \in N^-(i)} \xi_{ji}y_{ji}
 \right).
 \end{align}

We denote the set of all players in $N$ which are secured, i.e., $\sigma_i = 1$, under the above network-optimal security strategy by $\sStar$. We first observe that all players that opt to be secured under the independent security strategy continue to be secured under the network-optimal strategy. 

\begin{proposition}\label{prop_network}
Every player independently secured is also secured under the network-optimal security strategy, $\sStar \supseteq \sIndep$.
\end{proposition}

However, the positive externalities, inherent to this context, may result in certain nodes being secured under the network-optimal security strategy which are unsecured when acting independently. That is, we note that the above inclusion can be strict. We demonstrate this with \cref{ex:sInotSstar} in the e-companion. 

We now provide a key result demonstrating that the network-optimal security strategy and equivalently, $U(\graph)$, can be computed via a network-flow algorithm. The algorithm relies on the construction of an auxiliary directed network  $\graph^*$. We then establish a connection between the network-optimal security strategy in $\graph$ and the minimum weight $s$-$\ell$ cut problem in $\graph^*$. 

\subsubsection*{Construction of the Auxiliary Network ${\graph^*}$.} 
\begin{figure}[!htb]
  \begin{center}
    \includegraphics[scale = 0.175]{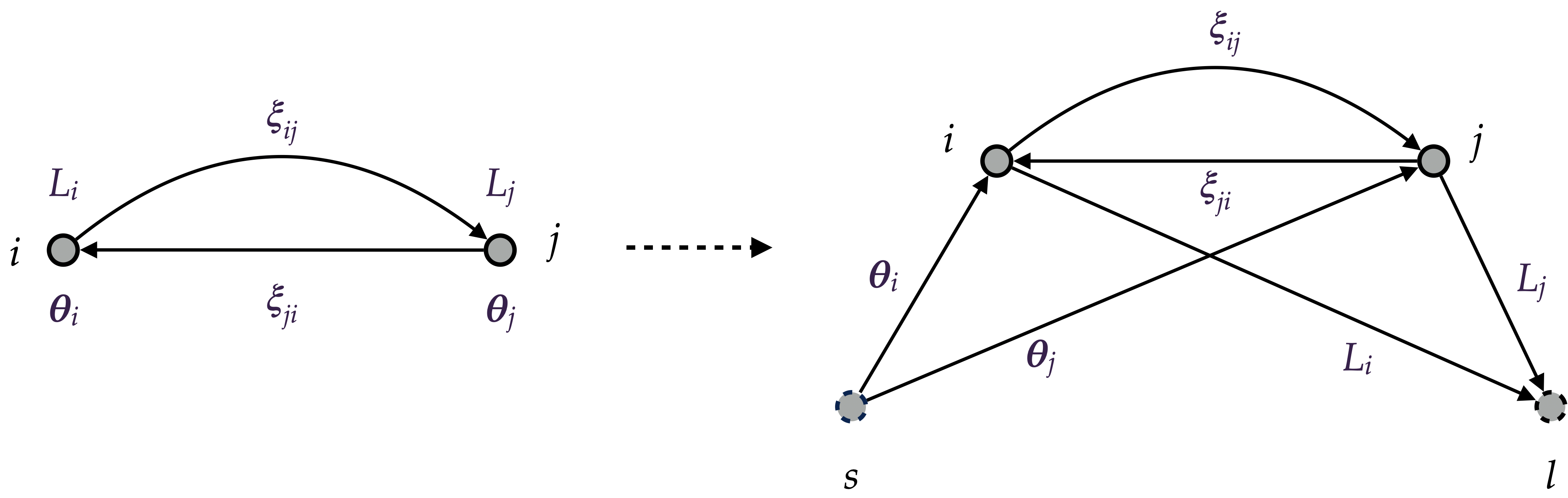}
    \caption{Auxiliary network $\graph^*$}
    \label{fig1}
  \end{center}
\end{figure}

The node set of $\graph^*$ is given by $N \cup \{ s, \ell\}$ where $s$ and $\ell$ are two additional nodes not present in the original network $\graph$. 
The nodes $s$ and $\ell$ represent the source and sink of the network $\graph^*$, respectively. 
The arc set of $\graph^*$ consists of, (i) arcs from $s$ to each node $i \in N$ with weights $\theta_i$, (ii) arcs from $i \in N$ to $j \in N^+(i)$ with weights $\xi_{ij}$,
(iii) arcs from $i \in N$ to $\ell$ with weights $L_i$. The construction of the auxiliary network is illustrated in \cref{fig1}. 

\begin{theorem}\label{thm:cutOptimal} 
Suppose the minimum weight cut separating $s$ and $\ell$ partitions the nodes of $\graph^*$ into $X$ and $\bar{X}$ such that $s \in X$. Then $\sStar = N\setminus X$. Further, $U(\graph)$ is the weight of the cut $(X, \bar{X})$. 
\end{theorem}
\smallskip

Also, from (\ref{eqn2}), it follows that if $\boldsymbol{x}^*$ and $\boldsymbol{y}^*$ denote the network-optimal security actions of the players, then, $x^*_i = 1$ if and only if $i \in \bar{X}$, and, $y_{ji} = 1$ if and only if $i\in X$, $j \in \bar{X}$. Therefore, from \cref{thm:cutOptimal}, we also immediately obtain the network-optimal security strategy. Now, note that the directed network $\graph^*$ has  $O(|N|)$ nodes and $O(|N|+|A|)$ arcs. Thus, from the push-relabel-algorithm \citep{goldberg1988}, we immediately obtain the following corollary. 

\begin{corollary}
$\sStar$ can be computed in $O((n^2+mn)log(n/m+n))$ time where $n = |N|$ and $m = |A|$.  
\end{corollary}

In the private information model, the network-optimal security strategy resolves two distinct kinds of inefficiencies engendered by the individually rational security strategies of the players. The first inefficiency arises from the canonical under-investment of efforts resulting from a failure to internalize positive externalities. This is well recognized in the interdependent security literature (see, for example, \citet{acemoglu2016}). Therefore, some agents for whom it was individually rational to not invest in security efforts are now secured since these erstwhile externalities are now internalized in the network-level optimization. This reflects the strategic complementarity inherent in situations with interdependent risks. The second source of inefficiency arises, in the private information model, as a consequence of security costs being privately held information. Equivalently, the non-inferability of security efforts of a player by other players who are not cooperating with it results in the inefficient duplication of security investments across the network. This provides an economic  rationale for anecdotal evidence from diverse supply chain security contexts that bear out this source of inefficiency \citep{asem2013}. 

Finally, we note the necessity of cost-sharing mechanisms in order to implement the network-optimal security strategy. For a player in the network, given the security states of all of its direct partner firms, the network-optimal security action is not necessarily individually rational. That is, the network-optimal security strategy is not always a Nash equilibrium strategy as demonstrated by \cref{exampleec1} provided in the e-companion.

\section{Security Cost Sharing Mechanisms}  
\medskip 

The next natural question is therefore to ask whether network-wide security cooperation in the private information model can be sustained with suitable cost-sharing mechanisms. Equivalently, we are interested in finding whether and when cooperation can be made individually rational, and the network-wide efficiency gains can be shared amongst the firms in a {\it stable} and {\it fair} manner. The field of cooperative game theory is well suited to address these questions. Towards that end, we first briefly review some cooperative game theory preliminaries. 

Cooperative game theory primarily addresses the question of whether cooperation can be sustained across a group of agents, and closely tied to this, is the problem of {\it fairly} sharing or allocation of profits (or cost savings) obtained via cooperation between those agents. A cooperative game is defined by $(N, c)$ where $N$ is the set of players in the game and $c(\cdot)$ is a characteristic function that associates to every subset (or, coalition) $S \subseteq N$ a corresponding cost $c(S)$. The subset consisting of all players, that is, the set $N$ itself is known as the grand coalition. An object of frequent interest is whether the grand coalition will form and whether it remains rational for individual players, or groups of players, to remain in the grand coalition.  In this work, we will only deal with cost games, i.e., where $c(S)$ is the cost incurred by coalition $S$, and players act to minimize their costs. A cooperative game $(N, c)$ is said to be {\it subadditive} if the characteristic function satisfies $c(S) + c(T) \geq c(S \cup T)$ for $S, T \subseteq N$.  Subadditivity can loosely be interpreted as offering an incentive for disjoint coalitions to cooperate. Another important property that a cooperative game can satisfy is convexity. The convexity property is stronger than the subadditivity property, and it loosely captures the intuition that as a coalition grows larger, the greater the incentive for other players to join it. Formally, $c(S) + c(T) \geq c(S \cup T) + c(S \cap T)$ for $S, T \subseteq N$.

\smallskip
\subsection{Interdependent Security Cost Sharing}
\smallskip 

Consider the set of agents $N$ situated on the graph $\graph$. Previously, the two security strategies considered represented the two extremes corresponding to no-cooperation and full-cooperation settings. We now extend the discussion to consider all intermediate levels of cooperation. That is, for any subset of agents, $S \subseteq N$, we define the {\it coalition-optimal security strategy} as that which minimizes the security cost of a cooperating set of agents $S$,  
  \begin{align}\label{coalitioncost}
    c(S)  &=\ \min_{\boldsymbol{x},\boldsymbol{y}}\quad  \sum_{i \in S} U_i(x_i, {\boldsymbol{y}_i}| I(i,S))   \nonumber \\ 
    &=\ \min_{\boldsymbol{x}, \boldsymbol{y}} \quad \sum\limits_{i \in S} \left( L_i(1-\sigma_i(x_i, {\boldsymbol{y}_i}| I(i,S))+ \theta_ix_i + \sum\limits_{j \in N^-(i)} \xi_{ji}y_{ji}\right).
  \end{align}

We define an indicator function $\security[S][i]$ for player $i$ belonging to a coalition $S$ that indicates whether player $i$ is secured under the coalition-optimal security strategy for $S$ in the private information model. Formally, $\security[S][i] = \sigma_i(\tilde{x}_i, {\boldsymbol{\tilde{y}}_i}| I(i,S))$, where $\tilde{x}_i$ and $\boldsymbol{\tilde{y}}_i$ denote the optimal solutions to (\ref{coalitioncost}). Further, denote the set of players secured in $S$ under the coalition-optimal security strategy by $\Upsilon(S)$. That is, $i \in \Upsilon(S)$ if and only if $\security[S][i] = 1$. Clearly, $S \backslash \Upsilon(S)$ are the players in $S$ that are not secured under the coalition-optimal security strategy. Further, for clarity, note that $\Upsilon(N) = \sStar$. The following result demonstrates a monotonicity property satisfied by the coalition-optimal security strategy that generalizes \cref{prop_network}. 

\begin{proposition}\label{monotone}\ A player $i \in S$ that is secured under the coalition-optimal security strategy for a coalition $S \subseteq N$ is also secured under the coalition-optimal security strategy for a coalition $T \supseteq S$, i.e., if $\security[S][i] = 1$, then $\security[T][i] = 1$. 
\end{proposition}
 
Further, the pair $(N, c)$ defines a cooperative game which we term as the {\it interdependent security cost sharing game}. This cost sharing game corresponds to our network model based on the private information assumption as clarified in \S2. In \S6, we will accordingly define and analyze the appropriate cost sharing game for the public information setting. 

The following proposition indicates that $c(S)$ can also be computed in polynomial time via a similar transformation to a minimum weight cut problem on the auxiliary graph $\graph^*$ as in \cref{thm:cutOptimal}.  

\begin{proposition}\label{coopgame_cost}
  $c(S)$ is the weight of the minimum cut separating the node set $N\backslash S$ and the node~$\ell$ in the auxiliary directed graph $\graph^*$ and thus can be computed in polynomial time. 
  \end{proposition}
\smallskip

An efficient security cost sharing mechanism is defined as $\boldsymbol{\phi}: (N,c) \rightarrow \mathbb{R}^n$ such that $\sum_{i\in N} \phi_i = c(N)$. An efficient security cost sharing mechanism is said to be a core allocation, i.e., it belongs to the core if and only if it is rational for all subsets of players in $N$ to remain in the grand coalition rather than deviate to form a coalition among themselves. That is, $\boldsymbol{\phi}$ is a core allocation if and only if, $\sum_{i\in S} \phi_i \leq c(S)\ \forall\ S \subseteq N$.  The core of some cooperative games may be empty. An empty core will preclude the existence of {\it stable} cost sharing arrangements. However, in cooperative games that are also convex, it is well known that the core of such games is non-empty \citep{shapley1971}. The following theorem demonstrating the convexity of the interdependent security cost sharing game therefore assumes significance since it guarantees the existence of a stable cost sharing mechanism. 

\begin{theorem}\label{convexity}
  The coalition-optimal security cost, $c(S)$, is submodular in $S$. Thus, the interdependent security cost sharing game $(N,c)$ always admits a stable security cost sharing mechanism.
\end{theorem}

Before we proceed to derive and analyze specific security cost sharing mechanisms, we observe that if a player is unsecured under the network-optimal security
strategy, then, the player is allocated $L_i$ by all stable cost sharing arrangements as formally demonstrated in \cref{restrict_assump}. Further, we also show that there exists a simple transformation of a network $\mathbb{G}$ where some players are unsecured under the network-optimal security strategy to another network $\mathbb{G}'$ where all players are secured in the network-optimal strategy and further, there exists a one-to-one correspondence between the core allocations of the
interdependent security games on $\mathbb{G}$ and $\mathbb{G}'$. Thus, \cref{restrict_assump}  allows us to restrict our attention to networks $\mathbb{G}$ and associated cost parameter vectors such that all firms are secured under the network-optimal security strategy.  

\subsubsection*{Shapley Value Based Security Cost Sharing.} The convexity of $(N, c)$ guarantees that a well-known and commonly employed allocation in cooperative games, the Shapley value \citep{shapley1953}, belongs to the core.  Beyond its membership in the core, the Shapley value also uniquely satisfies several natural fairness properties and has an axiomatic basis in general cooperative games. Formally, the Shapley value, $\Phi$, allocates to a player $i$ in a general cooperative game $(N,c)$,
\begin{eqnarray} \label {eq_shapley}
\Phi _i =\sum_{\{S:i \in
S\}}\frac{(|S|-1)!(n-|S|)!}{n!}\left(c(S)-c(S\setminus
\{i\}\right)).
\end{eqnarray}   

The Shapley value rewards players for their marginal contributions to various coalitions, and to that extent, it can be argued as exemplifying a certain notion of fairness. Further, $\Phi$ is the unique {\it efficient} allocation characterized by the following properties (or axioms):
\smallskip

\noindent \emph{i. Symmetry Property:} For players $i$ and $j$ such that for all subsets $S \subset N$, $i, j \notin S$, if $c(S \cup \{i\})-c(S) = c(S \cup \{j\})-c(S)$, then $\Phi_i=\Phi_j$.
\smallskip

\noindent \emph{ii. Null Player Property:} For player $i$ such that $c(S \cup
\{i\})=c(S)$ for all $S \subset N$, then $\Phi_i=0$.
\smallskip

\noindent \emph{iii. Additivity Property:} The Shapley value, $\Phi^{1,2}$, of a cooperative game, $(N, c^1+c^2)$, that is the sum of two cooperative games, $(N, c^1)$ and $(N, c^2)$, equals the sum of the Shapley values of the two games, $\Phi^1$ and $\Phi^2$, respectively.

Of these properties, we note that the symmetry property formalizes the idea that players which are ``identical'' in terms of their marginal contributions should receive an identical share of the value created by cooperation. This is, arguably, an innocent fairness criterion which, along with the marginal contribution interpretation discussed before, we shall return to later on in this work. The Shapley value is widely adopted as a cost-sharing or a profit-sharing, as the case may be, allocation method in
diverse contexts, including several mentioned in \S1.1, such as inventory pooling \citep{kemahliouglu2011}, capacity allocation and scheduling \citep{aydinliyim2010}, group purchasing \citep{chen2010}, disaster preparedness \citep{rodriguez2021}, and so forth. However, for our game, we establish a link between the computation of the Shapley value and the classical subset sum problem. In fact, this connection demonstrates that computing the Shapley value of interdependent security games is a computationally hard problem. 

\begin{theorem}\label{shapley_np}
There is no polynomial time algorithm that computes the Shapley value for a given player in the interdependent security cost sharing game unless P = NP.  
\end{theorem}

Further, from the proof of \cref{shapley_np}, we note that even for simple structures such as the assembly supply network, computing the Shapley value is hard. Beyond computational interest, the above result on the complexity of the Shapley value is of interest to us for reasons of implementation. In general, equilibrium concepts in non-cooperative game theory or solution concepts in cooperative games that are computationally intractable raise the question of feasibility of whether self-interested agents can identify and implement these mechanisms in practice.\footnote{Relatedly, \citet{roughgarden2010} observes, \textit{``(A) complexity-theoretic hardness result can diminish the predictive interpretation of an equilibrium concept and suggests more tractable alternatives [...] In a practical design context, it is obvious that a mechanism that is actually implemented had better be computationally tractable to run, like the deferred acceptance algorithm, and also easy to play, in the sense that participants should not need to perform difficult computations.''}}

For a notable special case, however, the Shapley value can be computed easily. In fact, when the expected penalties, in case of a realized risk, are sufficiently large for all players, then the Shapley value has a straight-forward closed form expression. 

\begin{theorem}\label{shapley_special}
If $L_i > \theta_i + \sum\limits_{j \in N^-(i)} \xi_{ji}$ for all $i\in N$, i.e., if  $\sIndep = \sStar = N$, then, the Shapley value based security cost allocation to player $i \in N$ is given by,
\begin{equation}
\Phi_i =  \theta_i + \sum_{j \in N^-(i)} \frac{\xi_{ji}}{2} - \sum_{j \in N^+(i)} \frac{\xi_{ij}}{2}.
\end{equation}
\end{theorem} 

In this scenario, when the expected penalties are sufficiently large, it is individually rational for all players to secure themselves (i.e., under the independent security strategy). That is, since all players choose to secure themselves even without cooperation, the network-optimal security strategy resolves only one kind of inefficiency, that arising from duplication of security efforts. Under the Shapley value based security cost sharing mechanism, in this scenario, the cost savings from avoiding duplication of security efforts across each link are equally shared by both parties. 

\subsubsection*{Extreme Core Allocations.} However, this still leaves open the question of whether, in general inter-firm networks, there exist stable security cost-sharing arrangements sustaining network-wide cooperation that can also be computed easily. We now provide an affirmative answer to this question. Consider an arbitrary permutation $\pi$ of the players in $N$. Then, we can define a cost-sharing allocation, $x_\pi$, corresponding to a permutation $\pi$ as follows, 
  $  x_{\pi_i} = c(\{\pi_1,\pi_2,\ldots,\pi_{i}\}) - c(\{\pi_1,\pi_2,\ldots,\pi_{i-1}\})\quad \forall i \in N.$

\begin{proposition}\label{extreme_core}
For every permutation $\pi$ of $N$, the allocation $x_\pi$ is an extreme point of the core of the interdependent security cost sharing game and can be computed in polynomial time.
\end{proposition}

The proof of \cref{extreme_core} relies on the convexity of the game and the characterization of the core of convex games as developed by \citet{shapley1971}. Further, we demonstrate that the extreme core points of the interdependent security cost sharing game can be computed in polynomial time, thereby, allowing us to identify easily computable and stable security cost sharing arrangements. However, it can easily be seen that extreme core allocations as identified in \cref{extreme_core} do not satisfy a basic notion of fairness as embodied in the symmetry property introduced earlier. 

\begin{proposition}\label{coresymmetry}
The security cost-sharing allocation $x_\pi$ does not satisfy the symmetry property.
\end{proposition}

Our discussion, thus far, uncovers what appears to be an ``impossible'' trilemma: {\it stability}, {\it fairness}, and {\it implementability}. That is, when we simultaneously require a security cost-sharing arrangement to be stable (i.e., it must be individually and coalitionally rational), fair (in terms of a basic symmetry property), and implementable (in terms of ease of computability), it already proves to be too restrictive. Descriptively, this suggests why, although the welfare gains achieved by network-wide security cooperation can, in principle, be stably shared, we may still not observe such cooperation in practice. In the next section, we will delve deeper into implementability concerns. Further, and importantly, we will also attempt to find a satisfactory reconciliation of the divergence between stability, fairness, and implementability. 

\section{Bilateral and Multilateral Implementability}  
\medskip

In \S4, we considered a narrow version of implementability. Specifically, we presumed a security cost-sharing mechanism that is easily computable is implementable.
However, implementing cost-sharing mechanisms via transfer payments across the network, even between firms that are not direct partners, is administratively challenging, perhaps even infeasible. 
Firms often have limited visibility let alone an ability to enter into cost-sharing arrangements with indirect network members. 
Therefore, in this section, we are prompted to study whether there exist stable and fair cost-sharing mechanisms that can be implemented via transfer payments only involving firms that are direct partners in the network. 
Indeed, since alliance networks are often comprised of a series of bilateral alliances in the first place, we develop a realistic bilateral implementation framework that can allow firms to sustain network-wide security cooperation against interdependent risks.\footnote{Furthermore, a purely cooperative-game theoretic approach to cost-sharing problems on occasion faces some criticism, as for example, in \citet{feng2021}, of providing \textit{"no implication for implementation in terms of how firms interact in the network and how financial payments are made among the firms."}}

To this end, we define the {\it bilateral implementability}  of a cost-sharing allocation as follows. A cost-sharing allocation $\Psi$ is bilaterally implementable if and only if for a given network $\mathbb{G}$ and associated cost parameter vectors $\{\boldsymbol{L}, \boldsymbol{\theta}, \boldsymbol{\xi}\}$, there exist differentiable functions $\{g_{ij}: j \in N(i)\}$ for each player $i \in N$ such that, 
\begin{align}\label{eqn7}
\Psi_i = \sum_{j \in N(i)} g_{ij}(\theta_i, \theta_j, L_i, L_j, \xi_{ij},
\xi_{ji}),
\end{align}
for cost parameters belonging to an open ball $\mathcal{B}^\epsilon$ centred at $(\boldsymbol{L}, \boldsymbol{\theta}, \boldsymbol{\xi})$ of radius $\epsilon$ for some $\epsilon > 0$. That is, qualitatively, the security cost apportioned to each player $i$ can be supported via verifiable transfer payments between only direct partners in the network. As discussed before, bilateral implementability obviates the need for transfer payments between firms not direct partners in the network. And consequently, since typically alliance networks expand via bilateral alliances, it also allows for sustaining network-wide cooperative security as the network structure evolves.

First, we examine the bilateral implementability of the Shapley value based security cost sharing allocation discussed in \S4. We introduce some definitions. For a given player $i \in N$, a set of players $P\subseteq N$ is said to be a {\it coalitionally rational security set} for $i$ if $i$ is secured in the coalitional optimal security strategy for the coalition $P\cup \{i\}$, i.e., $i \in \Upsilon(P\cup\{i\})$. We denote the set of all minimal\footnote{$P$ is said to be minimal if it is a coalitionally rational security set for $i$ but no subset of $P$ is.} coalitionally rational security sets for player $i$ by $\minSecG(i)$ and further, $\bar \minSecG(i) ={\textstyle \bigcup\limits_{P\in \minSecG(i)}}P$.

\begin{theorem}\label{shapley_bilateral}
Consider the Shapley value based security cost sharing allocation $\Phi$. \\
\noindent i) $\Phi$ is bilaterally implementable if for all players $i \in N$, $i \not \in \bar \minSecG(j)$ for all $j \in N(i)$ such that $|N(j)| > 1$.
\noindent ii) $\Phi$ is not bilaterally implementable if there exists a player $i \in N$ such that $i \in \bar \minSecG(j)$ for some $j \in N(i)$ such that $|N(j)\backslash N(i)| > 1$.
\end{theorem}

\cref{shapley_bilateral} provides characterizing conditions for when the Shapley value based cost sharing arrangement is bilaterally implementable. Observe that minimal coalitionally rational security sets formalize the externalities that secured players induce on other players in the network. Therefore, roughly speaking, the above theorem demonstrates that as the extent of positive externalities of security in the network increases, the Shapley value based security cost sharing fails to be bilaterally implementable. As a corollary, we observe that for the special case discussed in \cref{shapley_special}, the Shapley value cost-sharing mechanism is clearly bilaterally implementable. 

\cref{shapley_bilateral}, in conjunction with \cref{shapley_np}, arguably also demonstrates the impracticality of adopting a Shapley-value based security cost sharing arrangement in all but a narrow class of networks. Specifically, since it is neither computable efficiently nor bilaterally implementable, in general, we argue that this renders it contextually untenable. We now propose a novel security cost-sharing mechanism that builds on the extreme core allocations considered in  \cref{extreme_core}.  

\subsubsection*{Extreme Core Allocations and the Agreeable Allocation.} In light of \cref{restrict_assump}, we limit our attention to networks where all firms are secured in the grand coalition. We further recall the previously defined indicator function $\security[S][i]$ for player $i \in S$ that indicates whether player $i$ is secured under the coalition-optimal security strategy for $S$. That is, $\security[S][i] = \sigma_i(\tilde{x}_i, {\boldsymbol{\tilde{y}}_i}| I(i,S))$,
where $\tilde{x}_i$ and $\boldsymbol{\tilde{y}}_i$ denote the optimal solutions to (\ref{coalitioncost}). We now recursively define a finite family of mutually exclusive sets $\coreSubset = \{\coreSubset_1,\ldots,\coreSubset_\ell\}$ of players in the network where $\coreSubset_1 =\{i\in N:\security[\{i\}][i] = 1\}=\sIndep$. For $k > 1$, we define $\coreSubset_k$ recursively as,
\begin{align}
\coreSubset_k = \left\{
i \in N\setminus \bar{\coreSubset_{k-1}} :
\security[\bar{\coreSubset_{k-1}} \cup \{i\}][i] = 1
\right\},
\end{align} where $\bar{\coreSubset_{k-1}}=\coreSubset_1 \cup \ldots \cup \coreSubset_{k-1}$.
 In other words, $\coreSubset_1$ contains the players that are secured even under the independent security strategy, i.e., it is optimal for these players to secure themselves even when operating independently. 
 Further, $\coreSubset_2$ contains players that will be secured conditional on being in a coalition with players in $\coreSubset_1$, and so forth. 
 Also note that if $\coreSubset_k$ is a null set, then, so is $\coreSubset_{k+1}$. 
 Suppose there exists $\ell \in \mathbb Z$ such that $\bar{\coreSubset_{\ell}} = N$, then the recursive procedure generating the family of sets terminates. 
 Denote $s_k = |\bar{\coreSubset_k}|$ for $k = 1,\ldots,\ell$. 
 Then, any permutation $\pi$ of the players in $N$ such that $\pi_1,...,\pi_{s_1}$ is a permutation of players in $\coreSubset_1$, $\pi_{s_1 + 1},...,\pi_{s_2}$ is a permutation of players in $\coreSubset_2$, and so on up to, $\pi_{s_{\ell-1}+1},...,\pi_{s_\ell}$ is a permutation of players in $\coreSubset_\ell$ is defined as an {\em agreeable permutation}.

We note that it is possible in certain networks and associated cost parameter vectors for no $\ell \in \mathbb Z$ to exist such that $\bar{\coreSubset_{\ell}} = N$. In these cases, consequently, no agreeable permutation of the players in $N$ will exist either. Nevertheless, when the players in $N$ can be partitioned into the family of sets as described above, or equivalently, when an agreeable permutation of the players exists, we can demonstrate, as will be shown during the course of proving \cref{prop:BilateralSymmAlloc}, that the extreme core allocation $x_\pi$ corresponding to each agreeable permutation $\pi$ of $N$ is bilaterally implementable.

Furthermore, recall that extreme core allocations are not symmetric therefore, arguably, violating a basic notion of fairness. To remedy this, we are now in a position to propose our novel security cost sharing mechanism, the {\it agreeable allocation}, that is defined as the average of those extreme core allocations induced by all agreeable permutations of $N$.

\begin{theorem}\label{prop:BilateralSymmAlloc}
    The agreeable allocation of network-wide security costs, when it exists, (i) belongs to the core, and is,
    (ii) polynomial-time computable, 
  (iii) symmetric, and 
  (iv) bilaterally implementable. Further, it also satisfies,
  (v) marginality, and the
  (vi) null player property. Moreover, the security cost allocated to player $i$ by the agreeable allocation $x^*$ is given by, 
  \begin{align*}
  x^\star_i =\ \ \theta_i +  \sum_{\substack{j\in N^-(i)\\ j \in (N\setminus \bar{\coreSubset_k})}} \xi_{ji} - \sum_{\substack{j\in N^+(i) \\ j \in \bar{\coreSubset_{k-1}}}} \xi_{ij} + \sum_{\substack{j\in N^-(i) \\ j \in \coreSubset_k}} \frac{\xi_{ji}}{2} - \sum_{\substack{j\in N^+(i) \\ j \in \coreSubset_k}} \frac{\xi_{ij}}{2}   & \text { for }i \in \coreSubset_k.
\end{align*}
\end{theorem}
\medskip

Observe that the network-wide security cost apportioned to each player by the agreeable allocation depends only on its own security cost parameters and that of its direct partners, and therefore, it is bilaterally implementable. Also, importantly, we note that the agreeable allocation attempts to resolve the tension between {\it stability}, {\it fairness}, and {\it implementability}. Since, it belongs to the core, when it exists, it is a stable allocation of security costs. Further, in contrast to extreme core allocations, since it satisfies symmetry and marginality, it is in accordance with basic axiomatic descriptions of fairness. Further, in contrast to the Shapley value based cost sharing arrangement, since the agreeable allocation is computable in polynomial time, and saliently, is bilaterally implementable, it also fares well with respect to implementability concerns. Finally, the closed-form expression for the agreeable allocation provided above allows for transparency in the manner in which it allocates the network-wide security costs to each individual firm. In fact, the algorithm to compute the agreeable allocation and the closed-form expression lend themselves naturally to a straight-forward implementation mechanism.

We also remark that for the special case considered in \cref{shapley_special}, i.e., when $\sIndep=\sStar$, the agreeable allocation exists and coincides with the Shapley value.

\subsubsection*{Multilateral Implementability and  $\delta$-Agreeable Allocations.} 

The agreeable allocation is indeed appealing since its bilateral implementability minimizes the coordination challenges involved in sustaining the network-optimal security strategy. However, sometimes firms that are not direct partners may regardless cooperate via suitable transfer payments when it can be mutually beneficial. Consider a network $\mathbb{G}$ with associated cost parameter vectors $\{\boldsymbol{L}, \boldsymbol{\theta}, \boldsymbol{\xi}\}$. Formally, for an integer $\delta \geq 1$, a cost-sharing allocation $\Psi$ is said to be {\it $(\delta+1)$-laterally implementable} if and only if for cost parameters belonging to an open ball $\mathcal{B}^\epsilon$ centred at $(\boldsymbol{L}, \boldsymbol{\theta}, \boldsymbol{\xi})$ of radius $\epsilon$ for some $\epsilon > 0$, there exist differentiable functions $\{g_{ij}: j \in N, d(i,j) \leq \delta \}$ for each player $i \in N$ such that $\Psi_i = \Sigma_{j \in N, d(i,j) \leq \delta} \ g_{ij}$ where $g_{ij}$ is a function solely of the security cost parameters of players $i$ and $j$, and where $d(i,j)$ denotes the distance between nodes $i$ and $j$ in the network $\mathbb{G}$. That is, ($\delta+1$)-lateral implementability of a cost sharing allocation permits transfer payments between players that are at a distance of at most $\delta$ in the network. As $\delta$ increases, we expect the coordination challenges associated with the cost sharing mechanism to also increase. 

While our general approach to construct a $(\delta+1)$-laterally implementable allocation bears some resemblance to the previous development of the agreeable allocation, there are substantial technical differences. In the interest of brevity, we provide these details in the e-companion, \S EC.2. Broadly, we first identify a subset of permutations of the players in $N$ denoted as $\delta$-agreeable permutations (\cref{alg:rhoAgree}). A $\delta$-agreeable permutation can be computed via a fixed parameter tractable algorithm with respect to $\delta$ (i.e., polynomial time in $|N|$ but not in $\delta$). We then demonstrate that the extreme core allocations corresponding to each $\delta$-agreeable permutation is $(\delta+1)$-laterally implementable (\cref{prop:ec1}). We then define the $\delta$-agreeable allocation as the average of extreme core allocations induced by all $\delta$-agreeable permutations of $N$. 

\begin{theorem}\label{thm_deltaagreeable}
For a given integer $\delta \geq 1$, the $\delta$-agreeable allocation, when it exists, (i) belongs to the core, 
  (ii) is symmetric, and is,
  (iii) ($\delta$+1)-laterally implementable. Further, it also satisfies,
  (iv) marginality, and the
  (v) null player property.
\end{theorem}

The $\delta$-agreeable allocation satisfies the generalized notion of $(\delta+1)$-lateral implementability while retaining the fairness and stability properties of the agreeable allocation. Since the number of $\delta$-agreeable permutations can be exponential in $|N|$, the $\delta$-agreeable allocation is, in general, not computable in polynomial time for $\delta > 1$. However, as noted above, the $\delta$-agreeable allocation can be computed via a fixed parameter tractable algorithm, i.e., polynomial time in $|N|$ for a given $\delta$. In comparison, we note that the Shapley value allocation is also not, in general, computable in polynomial time but since it involves the consideration of all permutations of $N$ unlike the $\delta$-agreeable allocation which only considers a subset of permutations of players in $N$, the $\delta$-agreeable allocation is, in comparison, computationally less expensive, especially so when $|N|$ is large and $\delta$ is a fixed small number. In \S EC.2, we also provide \cref{example2} that clarifies the computation of the $\delta$-agreeable allocation and illustrates the notion of $(\delta+1)$-lateral implementability. 

\begin{theorem}\label{deltaagreeable_existence} Consider the interdependent security cost sharing game under private information.
    \begin{enumerate}[label=\roman*.] 
    \item If for an integer $\delta \geq 1$, the $\delta$-agreeable allocation exists, then the $(\delta+1)$-agreeable allocation also exists and coincides with the $\delta$-agreeable allocation. 
        \item For every integer $\delta \geq 1$, there exist networks $\mathbb{G}$ with corresponding security cost parameters such that the $\delta$-agreeable allocation does not exist but the $(\delta+1)$-agreeable allocation exists. 
    \item The $n$-agreeable allocation always exists where $n = |N|$. 
    \item The $n$-agreeable allocation coincides with the Shapley value allocation if and only if none of the $\delta$-agreeable allocations exist for $\delta < n$.
    \end{enumerate}
\end{theorem}

\cref{deltaagreeable_existence} clarifies a hierarchy of existence for $\delta$-agreeable allocations. As $\delta$ increases, and firms that are farther away from each other in the network are allowed to cooperate with each other via suitable transfer payments, the $\delta$-agreeable allocation is more likely to exist. However, naturally, as $\delta$ increases, arguably, the $\delta$-agreeable allocation becomes more challenging to implement than the agreeable allocation since it requires coordination between firms that are farther away in the network. Further, it follows from \cref{deltaagreeable_existence}(iv), and since in general, the Shapley value allocation involves transfer payments between any two firms in the network, $\delta$-agreeable allocations are (weakly) less challenging to implement than the Shapley value. 

\section{Network Security Model With Public Information}
\medskip

In this section, we consider the public information model, as presented in \S2, wherein all network cost parameters and actions are known to all players in the network. That is, the information set of every player $i$ in any coalition $S \subseteq N$ includes the security cost parameters and actions of all players in the network, $I(i,S) = \lbrace \theta_j, \xi_{kj}, L_j, x_j, y_{kj} : j \in N, k \in N^-(j)\rbrace$. Further, since a player can observe and infer the security actions of all other players in the network, player $i$ no longer needs to form a worst-case belief\footnote{In the general partial information model analyzed in \S EC.4, firm $i$ only adopts a worst-case belief for firms whose information is private, i.e., for $j \in N\setminus \mathcal{P}$, $\sigma_{ji} = 0$ whereas for $j \in \mathcal{P}$, $i$ forms an accurate belief, $\sigma_{ji} = \sigma_j$.} on the security state of other players $j \in N$, i.e, $\sigma_{ji} = \sigma_j$. And thus, firm $i$ ends up minimizing its its expected cost rather than its worst-case expected cost. 

Characterizing the security strategy of a coalition, or even the independent security strategy, in the public information model poses some challenges. In our network security model, as is often the case in network games with public information \citep{galeotti2010network}, there could be multiple Nash equilibria. Further, in the public information setting, the actions of a player or a coalition also depends on the actions of other players, and therefore, naturally on whether other players in the network are cooperating with each other. Therefore, we cannot analyze the security actions of a player or a coalition in isolation. We instead need to consider the cooperation structure across the entire network. This in contrast to the interdependent security cost sharing game developed in \S3 wherein the security cost of a coalition $S$ could be expressed independent of considering the actions of other players. Therefore, the interdependent security cost sharing problem under public information is modelled as a {\it cooperative game in partition function form} (see, e.g., \citet{hafalir2007efficiency}, \citet{fang2020}). Formally, given a partition $\rho$ of the players into disjoint coalitions whose union is $N$, the total security cost incurred by a coalition $S \in \rho$ in equilibrium is denoted by $\hat{c}(S; \rho)$. 

Again, we first consider the security actions of players when they are all acting independently. That is, $\rho$ consists of singleton sets of players. Each player $i \in N$ considers its security actions independently but knows all cost parameters in the network and can therefore infer the security actions of other players. Let $\hat{\Upsilon}^i_{\{i\}; \rho}$ be an indicator function denoting the equilibrium security state of player $i$ acting independently where $\rho$ is the coalition structure with all players in independent singleton coalitions. To address the multiplicity of equilibrium outcomes, we adopt a specific equilibrium selection procedure. Initially, all players choose their security actions independently without regard to the actions of other players in the network. Then, in subsequent rounds, players reassess their actions given the actions of others in preceding rounds. This procedure\footnote{Our equilibrium selection procedure bears resemblance and is motivated by the level-$k$ approach \citep{stahl1995players} which yields sufficient conditions for an equilibrium.} is formally described (\cref{alg:public_independent}) in the e-companion \S EC.3. Details and proofs for the results in this section are also provided in the e-supplement EC.3 in the interest of brevity.

\cref{alg:public_independent} computes an equilibrium security state of player $i$, $\hat{\Upsilon}^i_{\{i\}; \rho}$, in polynomial time. Given a general coalition structure $\rho$, we denote an equilibrium security state of player $i$ in coalition $S$ by $\hat{\Upsilon}^i_{S; \rho}$. The equilibrium selection procedure described above for the case of independent coalitions can similarly be extended (\cref{alg:public_coalition}) to compute, in polynomial time, an equilibrium security strategy for a coalition $S \subseteq N$ with a general partition $\rho$ of $N$ with $S \in \rho$. 

We then obtain the total security cost of a coalition $S$ belonging to a general coalition structure $\rho$ of $N$, $\hat{c}(S; \rho)$, as follows, 

\begin{equation}
\hat{c}(S; \rho) = \sum\limits_{i \in S} \left( L_i(1-\hat{\Upsilon}^i_{S; \rho})+ \theta_i\hat{\Upsilon}^i_{S; \rho} + \sum_{\substack{(j, i) \in A \\ \hat{\Upsilon}^i_{S; \rho} = 1, \hat{\Upsilon}^j_{T; \rho} = 0}} \xi_{ji}\right),
\end{equation}\label{eqn12}

where $S$ and $T$ are (possibly identical) coalitions in $\rho$ with $i \in S$ and $j \in T$. For clarity, we note that for the grand coalition structure $\rho^*$, i.e., when all players cooperate with each other, the total security cost under the public information and private information settings are equal, $\hat{c}(N; \rho^*) = c(N)$. This is since even under the private information setting all players in the grand coalition are aware of all security cost parameters in the network.  

We demonstrate that in the interdependent security cost sharing game under public information, $(N, \hat{c})$, the grand coalition is not necessarily stable. This is in contrast to our earlier result (\cref{convexity}) that there always exists a stable security cost sharing mechanism under the private information setting. This can be explained by two drivers. First, in the public information setting, one of the benefits of cooperative security, the benefit from additional information acquisition is removed. Thus, the benefits  from cooperative security in the public information setting are arguably lower.  Second, public information engenders free-riding since firms can now anticipate and observe the security actions of other firms in the network and benefit from the cooperation of other firms in the network without participating in the grand coalition and sharing security costs. Such free-rider issues have also been identified in other contexts to hinder cooperation and stability of the grand coalition in other partition function form games (see, e.g., \cite{yi1997stable}).

\begin{proposition}\label{public_core}
The grand coalition in the interdependent security cost sharing game under public information, $(N, \hat{c})$, is not, in general, stable to defections. 
\end{proposition}

We now, however, show that the agreeable allocation can be extended to the public information setting while retaining several of its desirable properties. Notably, we prove that, analogous to \cref{prop:BilateralSymmAlloc}, the public information version of the agreeable allocation, when it exists, satisfies individual rationality, a weaker notion of stability wherein each player is better off in the grand coalition (i.e., with full cooperation) as compared to the independent coalitions (i.e., no-cooperation) scenario. 

\subsubsection*{Agreeable Allocation with Public Information.}  Again, for ease of exposition, we restrict our attention to networks where all firms are secured in the grand coalition. We recursively define a finite family of mutually exclusive sets $\mathcal{T} = \{\mathcal{T}_1,\ldots,\mathcal{T}_\ell\}$ of players in the network where $\mathcal{T}_1 =\{i\in N:\hat{\Upsilon}^i_{\{i\}; \rho_1} = 1\}$ where $\rho_1$ corresponds to the independent coalition structure. For $k \geq 1$, we then define $\mathcal{T}_{2k}$ and $\mathcal{T}_{2k+1}$ recursively as follows, where $\bar{\mathcal{T}_{k}}=\mathcal{T}_1 \cup \ldots \cup \mathcal{T}_{k}$. Further, the coalition structure $\rho_{k+1}$ contains the coalition $\bar{\mathcal{T}_{k}}$ and all other players in $N \setminus \bar{\mathcal{T}_{k}}$ are in independent coalitions. Also, recall that $\hat{\Upsilon}^i_{S; \rho}$ is the equilibrium security state of player $i \in S$ with the coalition structure $\rho$ in the public information model whereas ${\Upsilon}^i_{S}$ is the coalition-optimal security state of $i \in S$ in the private information setting. 
\begin{align}
\mathcal{T}_{2k} &= \left\{
i \in N\setminus \bar{\mathcal{T}_{2k-1}} :
\Upsilon_{\bar{\mathcal{T}_{2k-1}} \cup \{i\}}^{i} = 1\right\}\\
\mathcal{T}_{2k+1} &= \left\{
i \in N\setminus \bar{\mathcal{T}_{2k}} :
\hat{\Upsilon}_{\bar{\mathcal{T}_{2k}} \cup \{i\}; \rho_{2k+1}}^{i} = 1
\right\}\label{eqn14}
\end{align} 

$\mathcal{T}_1$ contains players that are secured under the independent coalition structure. That is, in the equilibrium outcome obtained from \cref{alg:public_independent}, these players are secured. $\mathcal{T}_2$ contains players who, if they are secured, save the costs of extrinsic security for players in $\mathcal{T}_1$ and bestow a direct positive externality to the players in $\mathcal{T}_1$ that outweighs their own cost of security. Thus, for the players in $\mathcal{T}_1 \cup \mathcal{T}_2$, it is optimal in the private information model as well to secure themselves. Further, there will be players in $\mathcal{T}_3$ for whom it is individually rational to secure themselves conditional upon players in $\mathcal{T}_1$ and $\mathcal{T}_2$ being in a coalition together, $\mathcal{T}_1 \cup \mathcal{T}_2$. Successive sets of players are identified iteratively. Note that these families of sets are constructed in a very similar manner as in the private information model. The only distinction arises in (\ref{eqn14}) from observing that in a public information model, the formation of each new coalition may also trigger a change in the security actions of other players who can respond to this. 

Suppose there exists $\ell \in \mathbb Z$ such that $\bar{\mathcal{T}_{\ell}} = N$, then the recursive procedure generating the family of sets terminates. Again, it is possible in certain networks and associated cost parameter vectors for no $\ell \in \mathbb Z$ to exist such that $\bar{\mathcal{T}_{\ell}} = N$. In these cases, consequently, no agreeable allocation will exist. Unlike in the private information setting where a closed form expression for the agreeable allocation is derived, the agreeable allocation under public information $\hat{x}$ is obtained by \cref{alg:publicAgreeable} provided in \S EC.3 which takes in the family of sets $\mathcal{T}$ as an input.

\begin{theorem}\label{thm:public_main} The agreeable allocation under public information, $\hat{x}$, computed by \cref{alg:publicAgreeable}, when it exists, is (i) individually rational, (ii) polynomial-time computable, and (iii) bilaterally implementable. Further, it also satisfies, (iv) symmetry, and the (v) null player property. 
\end{theorem}

Therefore, while the agreeable allocation cannot guarantee that the grand coalition is stable to defections by subsets of players (indeed no cost sharing allocation can), it still satisfies a weaker notion of stability. It ensures that all players will prefer to remain in the grand coalition structure $\rho^*$ rather than in the independent coalition structure. Further, we interestingly find that the public information version of the agreeable allocation exists if and only if the agreeable allocation as defined in the private information setting exists.  

\begin{corollary}\label{agreeable_existence}
For a given network $\mathbb{G} = (N,A)$ and associated security cost parameters, the agreeable allocation under public information $\hat{x}$ exists if and only if the agreeable allocation under private information $x^*$ exists. 
\end{corollary}

Here, we briefly comment on some main implications of our analysis of the general partial information model in \S EC.4. First, we demonstrate that the agreeable allocation can be naturally extended to the partial information model thereby generalizing \cref{thm:public_main}. Therein, we observe that \cref{agreeable_existence} also generalizes and the existence of the agreeable allocation is not contingent on the informational assumption in the network. Finally, and importantly, we clarify that even in the presence of partial public information in the network, the grand coalition may be unstable and that if the grand coalition is unstable with a certain level of public information in the network, it remains unstable at higher levels of information provisioning in the network. 

\section{Quasi-Homogeneous Networks}
\medskip

The chief deficiency of the agreeable allocation, under all informational assumptions is that, in general, depending on the structure of the interfirm network, or the associated security costs, it may not exist. To the extent that an agreeable allocation is viewed as desirable for its fairness, bilateral implementability, and other properties as documented in \cref{prop:BilateralSymmAlloc} and \cref{thm:public_main}, this offers a rationale for when inter-firm networks will find it challenging to cooperatively secure themselves. In order to examine the role of the network structure on the existence of the agreeable allocation, we now consider {\it quasi-homogeneous} networks $\mathbb{G}$ as networks wherein the costs of securing against intrinsic risks for firm $i$, $\theta_i$, are identical for all firms. Similarly, we also assume costs of securing against extrinsic risks, $\xi_{ij}$, are identical across all links in the network, and the expected penalties faced by players in the event of a realized risk are also equal. Formally, a network $\mathbb{G}$ is said to be quasi-homogeneous if $\theta_i = \theta$ and $L_i = L$ for all $i \in V$, and, $\xi_{ij} = \xi$ for all $(i,j) \in A$. 

Analyzing quasi-homogeneous networks permits us to isolate the effects of the network structure on the existence of the agreeable allocation. A priori, it is qualitatively unclear what the role of network structure would be on the existence of the bilaterally implementable agreeable allocation. For instance, denser networks can render it easier for efficient and stable cost sharing arrangements to be bilaterally implementable since there are more bilateral links. However, denser networks may also result in wider positive externalities to securing oneself necessitating multilateral cooperation.

We now introduce some graph-theoretic definitions that aid us in identifying when quasi-homogeneous networks admit and do not admit an agreeable allocation of security costs. We define a {\it k-core} of network $\graph$ as an induced subgraph $\mathbb{H}$ of $\graph$ such that the in-degree of all nodes in $\mathbb{H}$ is at least $k$.\footnote{Conventionally, $k$-cores are defined on undirected graphs. Herein, we consider a natural analogue for directed graphs.} Then, a $(k,\ell)${\it-core} is a $k$-core $\mathbb{H}$ of $\graph$ such that, if $\ell$ denotes the maximum out-degree of a node in $\mathbb{H}$ to the nodes in $\graph\backslash \mathbb{H}$, then $k > \ell$. 
Therefore, while a $k$-core is a sufficiently dense induced subgraph, a $(k,\ell)${\it-core} is an induced subgraph that is sufficiently dense internally and simultaneously sparse in its connections with other nodes in the graph. 

\begin{theorem}\label{agreeable_homogeneous}
Consider a quasi-homogeneous network $\graph$ with security cost parameters given by $L,\ \theta,$ and $\xi$. \\
\noindent i. $\graph$ admits an agreeable allocation if $\graph$ does not contain a k-core where $k =$ {\footnotesize $\left\lceil \frac{L-\theta}{\xi} \right\rceil$}.\\
\noindent ii. $\graph$ does not admit an agreeable allocation if $\graph$ contains a $(k,l)${\it-core} where $k = \ell +$ {\footnotesize $\left\lceil \frac{L-\theta}{\xi} \right\rceil$}.
\end{theorem}

The two parts of \cref{agreeable_homogeneous} provide distinct sufficient and necessary conditions, respectively, for the existence of the agreeable allocation in quasi-homogeneous networks. From a descriptive standpoint, it implies qualitatively that the agreeable allocation is guaranteed to exist in (quasi-homogeneous) networks so long as they are not sufficiently locally dense. This refines our earlier intuition on the role of interfirm network structure on the existence of the agreeable allocation. Further, in graphs that contain sufficiently dense and sufficiently local clusters, the agreeable allocation is guaranteed to not exist. 

\section{Numerical Case Study}\label{subsec_numerical}
\medskip 

\begin{figure}[!htb]
  \begin{center}
    \includegraphics[scale = 0.23]{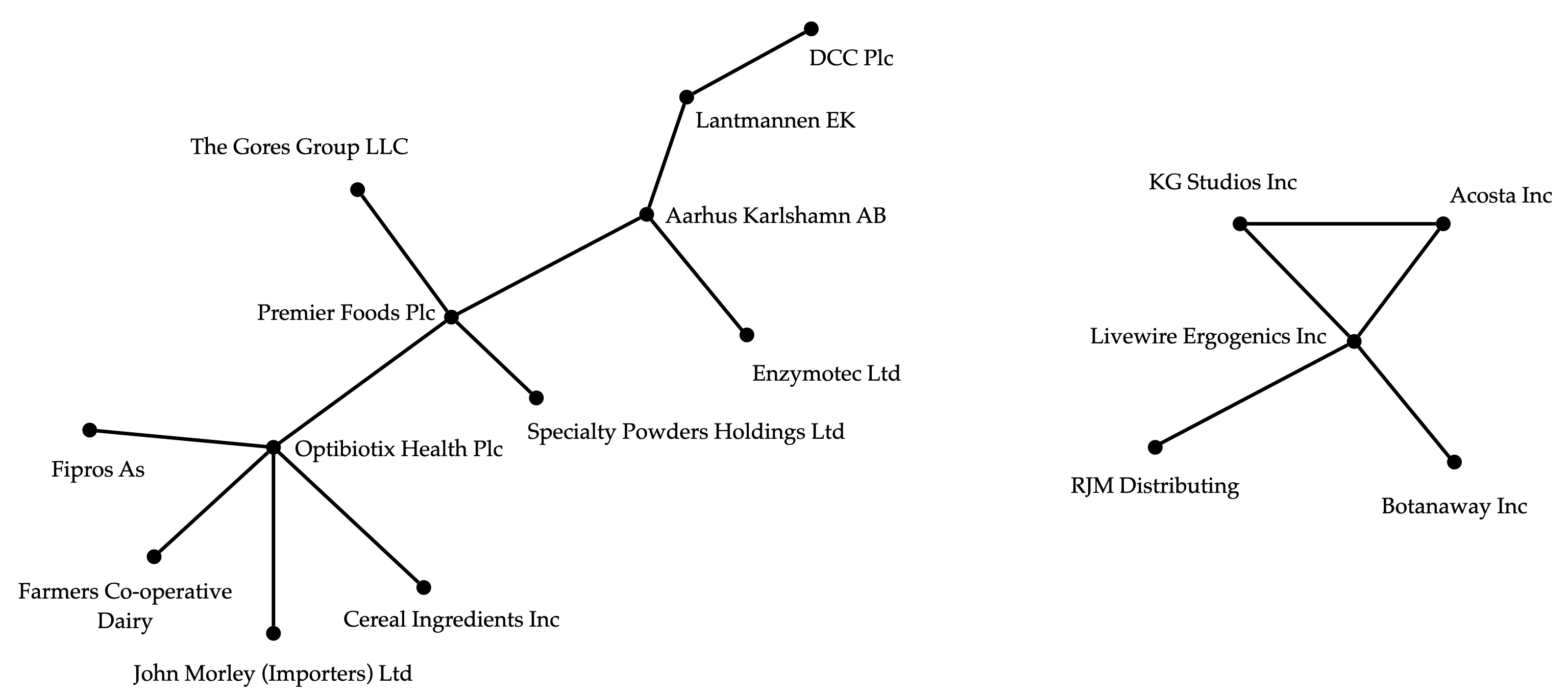}
    \caption{Examples of alliance networks in the food manufacturing sector.}
    \label{fig_networks}
  \end{center}
\end{figure}

We now present a case study analyzing the feasibility of cost sharing mechanisms to sustain network-wide cooperative security in real-world interfirm networks that can face interdependent risks. Specifically, we use the Refinitiv SDC Alliance database to extract all alliances in the food manufacturing sector formed between 2006 to 2020. 
The database contains 2339 alliances formed between 3073 unique firms in our industry of interest. Typically, these are bilateral alliances formed between two firms, while, on occasion, alliances are formed between three or more firms. For example, one of the alliances in the database is between Optibiotix Health Plc, a biotechnology company that manufactures SlimBiome, a weight management supplement, and John Morley (Importers) Ltd, which manufactures prepared perishable foods. Optibiotix Health would supply the weight management supplement to be included in prepared muesli packs manufactured by John Morley Ltd within the UK. In this example, the presence of an interdependent risk is evident. Over time, larger networks of alliances arise and we identify 792 distinct interfirm networks. Of these, the largest connected network of firms contains 1092 nodes. The other networks are smaller, and we remove all networks consisting of only two firms since these networks trivially permit bilaterally implementable cost sharing mechanisms. We in fact restrict our attention to alliance networks that are of size at least five and we obtain exactly 50 such alliance networks.\footnote{The largest alliance network comprises 1092 firms and 2624 partnerships (i.e., arcs). The other 49 alliance networks are smaller and qualitatively bear structural similarities containing an average of 6.79 nodes (a median of 6 nodes) and 14.28 arcs (a median of 12). The average degree of each node across the 50 alliance networks (i.e., the average number of partners for a firm) is 2.13. We also observe that 28 of these 50 alliance networks are {\it trees}.} We depict two of these networks in \cref{fig_networks}. 

We leverage the algorithmic results obtained in previous sections to numerically test whether the agreeable allocation exists, and when it exists, compute the network-wide security cost apportioned by the allocation. These results are meant to be illustrative since the existence of the agreeable allocation naturally depends on the precise security cost parameter specifications. However, the security cost parameters and the penalties are simulated in a systematic manner. Across all simulated networks, we set the parameter $\theta_i \sim U[15,25]$ for all  firms $i$, and for all links between firms $i$ and $j$, $(i,j)$, $\xi_{ij} \sim U[3,5]$. Further, for all $i$, $L_i \sim U[17+\delta_i, 23 + \delta_i]$, where $\delta_i = |N^-|$. That is, we assume that firms with more partners are larger firms and thus, also likely to incur higher reputation costs. Based on 1000 simulated runs for each of the 50 alliance networks, we make the following observations. 

First, we observe that in 56.7\% of the simulated networks, the agreeable allocation exists. In contrast, in only 0.79\% of the simulated networks, the Shapley value based security cost sharing allocation is of the form given by \cref{shapley_special} and hence, bilaterally implementable. This, in conjunction with the straight-forward implementation mechanism described in \S5, demonstrates the practical relevance of our proposed security cost sharing allocation. Second, we find, interestingly, that the alliance network permitting the agreeable allocation to exist with the highest likelihood of 74.3\%, is a star network. Finally, we observe that the networks which rarely permit the existence of the agreeable allocation, in only 2.6\% and 4\% of the simulations, respectively, are both completely connected networks, i.e., cliques of size six. This lends further evidence in support of \cref{agreeable_homogeneous} that densely connected networks preclude the existence of the agreeable allocation. 

In the above numerical experiment, the cost parameters for all nodes in a network were drawn from the same distributions. However, in real-world networks, there is usually a significant asymmetry in the penalties incurred by firms in case of a realized risk. Consumer-facing firms typically incur substantially larger penalties than others. To incorporate this in our simulation, we obtain the Standard Industrial Classification (SIC) codes of the firms from the SDC database. We then denote firms in the retail industry (with an SIC code in the range 5200 to 5999) as consumer-facing firms. Of the 3073 unique firms in our dataset, we identify 154 such (potentially) consumer-facing firms. In our second numerical experiment, we simulate the cost parameter $L_i$ for a firm $i$ such that a consumer-facing firm faces a larger expected penalty and the expected penalty decays exponentially with the distance from the consumer, i.e., $L_i = L_0/c_0^{d_i}$, where $d_i = 0$ if $i$ is a consumer-facing firm, $L_0$ is the expected penalty it faces and $c_0$ is a constant.\footnote{$L_0$ was chosen to be 409.6 (=$2^{11}/5$) and $c_0= 2$ for the results reported here.}  We again perform 1000 simulation runs for each of the 50 alliance networks. Each network is then compared against a benchmark simulation wherein the penalties of all firms are drawn from the same uniform distribution with an expected penalty given by $\left(\sum\limits_{i \in N} L_0/c_0^{d_i}\right)/N$. This allows us to comment on the role of cost asymmetry on the existence of the agreeable allocation vis-\'{a}-vis the bilateral implementability of the Shapley mechanism. For our chosen parameter values, we find that in the benchmark network simulations, the Shapley value nearly always coincides with the agreeable allocation and is bilaterally implementable for all of the 50 networks. However, with asymmetric penalties, the Shapley value is bilaterally implementable only in $34.47$\% of the simulations. For 15 of the 50 networks, it was never bilaterally implementable across all 1000 runs. In contrast, the bilaterally implementable agreeable allocation exists in $71.35$\% of the simulated networks. Across various choices of $L_0$ and $c_0$, we recover qualitatively identical results. In summary, in real-world networks with cost asymmetries, despite the non-existence of the agreeable allocation in certain instances, the practical advantage\footnote{Not surprisingly, we also observed a substantial advantage in terms of the computational time required to obtain the agreeable allocation in comparison to the Shapley value.} of the agreeable allocation in terms of its bilaterally implementability over the Shapley mechanism is further underscored. 

\section{Concluding Remarks}
\medskip

Networked firms are exposed to a variety of interdependent, or contagion, risks such as supply chain contamination, deliberate adulteration, or cybersecurity threats and data breaches. The fundamental distinction that sets apart these risks from other types of risks faced by firms is their transferable nature. In this paper, we develop a network model to study the cooperative management of interdependent risks by networked firms. 
\smallskip

The network-wide cooperative security strategy in our interdependent risk model can be computed in polynomial time via a minimum-weight cut network flow algorithm. Assuming that the security costs and actions are private information known only to the respective players, we find that firms have a clear incentive to cooperate and that there exist stable security cost-sharing mechanisms that can sustain network-wide cooperation. However, in the presence of public information, we find that, in general, there do not exist cost-sharing mechanisms that can ensure the stability of the grand coalition. Thus, it appears that interdependence of network security is alone insufficient to sustain network-wide cooperation.  

Introducing the notion of bilateral implementability, we uncover a fundamental trilemma between stability, fairness, and implementability of network security cost-sharing mechanisms. We then develop a novel cost sharing mechanism, the {\it agreeable allocation}, which attempts to balance the three notions. The agreeable allocation, when it exists, satisfies notions of stability, is formalizably fair, easily computable, and is also implementable via a series of bilateral cost sharing agreements. However, the agreeable allocation may not always exist. This, we argue, once again, demonstrates that, although cost-sharing mechanisms belonging to the core can be identified, sustaining network-wide security cooperation can still be challenging and therefore, may not always be possible in practice. We then construct $\delta$-agreeable allocations that satisfy the general notion of $(\delta+1)$-implementability which permits firms that are not direct partners to also enter into cost-sharing agreements if they are at a distance of at most $\delta$ from each other in the network. As $\delta$ increases, the $\delta$-agreeable allocation is more likely to exist. However, as $\delta$ increases, we also expect the coordination challenges to increase thereby highlighting a fundamental trade-off.

Moreover, to study the role of network structure on the existence of the agreeable allocation, we consider quasi-homogeneous networks (i.e., networks with homogeneous costs of security and expected penalties in case of realized risk), and find that networks without sufficiently dense clusters admit an agreeable allocation. Whereas, networks containing sufficiently dense and local clusters do not permit an agreeable allocation of network-wide security costs. Finally, using the SDC alliance database, we extract all alliances formed in the food manufacturing sector between 2006 to 2020. With numerical experiments and simulated cost parameters, we argue the practical feasibility and relevance of employing the agreeable allocation as a bilateral security cost-sharing mechanism in real-world alliances to sustain network-wide cooperative security against interdependent risks. 
\smallskip

This work develops, to the best of our knowledge, for the first time, an economic theory of cooperative security against interdependent risks in networks. However, we acknowledge several limitations and open problems arising from our study. 
\smallskip

\noindent\textbf{Limitations.} Certainly, there are some important questions that remain to be answered. First, for instance, the question of the general existence (or non-existence) of a bilaterally implementable and stable cost sharing mechanism remains open. Second, and crucially, in this paper, we consider interfirm networks characterized by repeated and ongoing interactions between firms. Thus, a vulnerable firm is nearly certain to transfer risks to its partner firms if the partner firms do not secure the corresponding link. A richer model of interdependent security would allow for a stochastic transmission and propagation of risk in the network. However, this richer stochastic model of interdependent network security is challenging to analyze. Particularly, the characterization of cooperative security strategies in this stochastic model of interdependent security is a non-trivial problem. Finally, we assume that the considered networks are static whereas, in reality, networks tend to change dynamically, with new alliances being formed, and existing alliances being broken over time. Bilaterally implementable cost-sharing mechanisms, in particular, may be well-suited to sustain cooperation in dynamic alliances, as we have noted earlier. 

\medskip

\bibliographystyle{plainnat}

\bibliography{ref}

\newpage
\appendix
\begin{center}
\noindent{\large{{\bf Electronic Companion: Cooperative Security Against Interdependent Risks}}}\label{appendix}
\end{center}
\medskip

\section{Proofs and Technical Results}
\medskip 

\noindent\textbf{Proof of \cref{prop_independent}.} Consider a player $i \in N$. First, note that under the independent security strategy, the worst-case security state of player $i$ as a function of its information set and security actions is given by, 

\begin{equation*}
\sigma_i(x_i, {\boldsymbol{y}_i} \vert I(i,\{i\})) =  
\begin{cases}
    0, &\text{ if } x_i \left(\prod_{j \neq i} y_{ji}\right) = 0,\\ \\
    1, &\text{ otherwise.}\\
  \end{cases}
\end{equation*}
\smallskip

Therefore, $\sigma_i = 1$ if and only if $x_i = y_{ji} = 1$ for all $j \in N^-(i)$. Further, if $\sigma_i = 0$, then $U_i$ is minimized when $x_i = y_{ji} = 0$ for all $j \in N^-(i)$. We now analyze these two cases in succession. If $\sigma_i = 0$, then the minimum worst-case expected cost $U_i = L_i$. If $\sigma_i = 1$, then $U_i = \theta_i + \sum\limits_{j : (j,i) \in A} \xi_{ji}$. Therefore, player $i$ is independently secured, i.e., belongs to $\sIndep$ when $U_i$ is minimized at $\sigma_i = 1$. That is, $i \in \sIndep$ if and only if $\theta_i + \sum\limits_{j : (j,i) \in A} \xi_{ji} \leq L_i$. \hfill $\square$
\bigskip 

\noindent\textbf{Proof of \cref{prop_network}.} Suppose that player $k \in N$ is secured under the independent security strategy. We will now show that $k$ will remain secured under the network-optimal security strategy. Consider $U(\graph)$ and let $x^*_i$, $y^*_{ji}$ for $j\in N^-(i)$ denote the network-optimal security actions by any player $i \in N$. Suppose, to the contrary, that $k$ is unsecured in the network-optimal security strategy, that is, $x^*_k = y^*_{jk} = 0$ for $j\in N^-(k)$. Consider an alternate security strategy such that $x_i = x^*_i$ and $y_{ji} = y^*_{ji}$ for all $i \neq k$ and $j \in N^-(i)$, and $x_k = y_{jk} = 1$ for $j\in N^-(k)$. Then, it is clear from (\ref{eqn2}) that the security state of every player remains the same except for $k$ who is now secured under the new security strategy. Therefore, 
\begin{align*}
\sum_{i \in N} U_i(x_i, {\boldsymbol{y}_i}| I(i,N)) &= 
 \theta_i + \sum\limits_{j : (j,k) \in A} \xi_{jk} + \sum_{i \neq k} 
 \left( 
    L_i(1-\sigma_i(x^*_i, {\boldsymbol{y}^*_i}| I(i,N))) 
    + \theta_ix_i 
    + \sum_{j : (j,i) \in A} \xi_{ji}y_{ji}
 \right) \\
 &\leq L_i + \sum_{i \neq k} 
 \left( 
    L_i(1-\sigma_i(x^*_i, {\boldsymbol{y}^*_i}| I(i,N))) 
    + \theta_ix_i 
    + \sum_{j : (j,i) \in A} \xi_{ji}y_{ji} \right ) \\
    &= U(\graph).
 \end{align*}
 
 The inequality follows from \cref{prop_independent} yielding a contradiction to the minimality of $U(\graph)$. Therefore, $k$ has to remain secured under the network-optimal security strategy. Consequently, $\sIndep \subseteq \sStar$. \hfill $\square$

\noindent\begin{example}[$\sIndep \subset \sStar$ strictly]
\label{ex:sInotSstar}
Consider a set of players, $N = \{1, 2, 3\}$ situated on a graph with arc set $A =\{(2,1), (2,3)\}$. Further, let $\theta_i = 0,\, L_i = 1$ for all $ i \in N $, and let $\xi_{21} = \xi_{23} = 2$. It can be easily verified that $\sIndep = \{2\}$, however, $\sStar = \{1, 2, 3\}$.
 \hfill \Halmos
 \end{example}
\bigskip 

\noindent\textbf{Proof of \cref{thm:cutOptimal}.}

Let $\sigma^*_i$ denote the {network-optimal} security state of player $i$, i.e., $\sigma_i^* = 1$ if and only if $i\in\sStar$.
Further, let us denote by $x^*_i$, and $y^*_{ji}$ for $j\in N^-(i)$, the {network-optimal} security actions by player $i \in N$. 
We first note that {for $i \in \sStar$} from (\ref{eqn2}) with $S = N$, $x^*_i = 1$, and $y^*_{ji} = 1$ for all $j \in N\backslash \sStar$. 
Further, $y^*_{ji} = 0$ for all $i, j \in \sStar$ since if players $i$ and $j$ are both secured, it is not optimal to secure the links between them. 
Moreover, for $i \in  N\backslash \sStar$, that is when $i$ is unsecured, it is not optimal for $i$ to partially secure itself from intrinsic or extrinsic risks. 
Therefore, $x^*_i = 0$, and $y^*_{ji} = 0$ for all $j \in N\backslash \{i\}$. 
Then, 
\begin{align*}
U(\graph) &=\ \sum_{i \in N} U_i(x^*_i, {\boldsymbol{y}^*_i}| I(i,N))   \nonumber \\
 &=\ 
 \sum_{i \in N} 
 \left( 
    L_i(1-\sigma_i(x^*_i, {\boldsymbol{y}^*_i}| I(i,N))) 
    + \theta_ix^*_i 
    + \sum_{(j,i) \in A} \xi_{ji}y^*_{ji}
 \right) \\
 &=\ \sum_{i \in \sStar} 
 \left( \theta_i + \sum_{j \in N\backslash \sStar, (j,i) \in A} \xi_{ji} \right) +  \sum_{i \in N\backslash \sStar} L_i.
 \end{align*}

 Now, consider the auxiliary network $\graph^*$ and the minimum weight directed cut $(X, \bar{X})$ separating $s$ and $\ell$ in $\graph^*$ with source $s \in X$ and sink $\ell \in \bar{X}$. 
 The minimum weight directed cut in this network identifies $X$ and $\bar{X}$ such that the sum of weights on arcs directed from $X$ to $\bar{X}$ is minimized. 
 The sum of weights of these arcs is given by, 
 \begin{align*}
w(X,\bar{X}) = \sum_{i \in \bar{X}} \left(\theta_i +  \sum_{j \in X, (j,i) \in A} \xi_{ji} \right) +  \sum_{i \in X} L_i.  \nonumber 
 \end{align*}
Comparing the expressions, $w(X,\bar{X})$ and $U(\graph)$ are simultaneously minimized when $\bar{X} = \sStar$ and $X = N\backslash \sStar$. This completes the proof. 
\hfill $\square$
\medskip

The following example demonstrates that the network-optimal security actions are not always individually rational for the players. Thus, cost-sharing mechanisms are required for firms to adopt and sustain the network-optimal security strategy.

\begin{example}\label{exampleec1}
Let $N=\{1,2\},\,A=\{(1,2), (2,1)\}$. Let $\theta_1 = \theta_2 = 2$, $L_1 = 6, L_2 = 1$. Further, let $\xi_{12} = 1,\, \xi_{21} = 2$. It is easily seen that the network-optimal security strategy secures both players. However, even {\em given} that $1$ is secured, it is still not individually rational for $2$ to secure itself since its expected penalty is lower than its instrinsic security cost. Thus, the network-optimal security strategy is not a Nash equilibrium strategy. This example demonstrates that in order to implement and sustain the network-optimal security strategy, transfer payments between the players are necessary. 
 \hfill \Halmos
\end{example}
\medskip 

\noindent\textbf{Proof of \cref{monotone}.} Suppose $S \subset T$ and let $\Upsilon(S)$, $\Upsilon(T)$ denote the set of secured players under the coalition-optimal security strategies of coalitions $S$ and $T$, respectively. Then, let $X$ denote $\Upsilon(S)\backslash\Upsilon(T)$, $Y = \Upsilon(S)\cap\Upsilon(T)$, and $Z = \Upsilon(T)\backslash \Upsilon(S)$. Then, if $X$ is an empty set, then our proof is complete, since, then $\Upsilon(S) \subseteq \Upsilon(T)$. Therefore, suppose $X$ is not an empty set. Then, consider the change in the coalition-optimal security cost $c(T)$ if the nodes in $X$ were also secured. The change in the coalition-optimal security cost will be given by, $\theta(X) - L(X) + \xi(N\backslash(X\cup Y\cup Z) - \xi(X, Y \cup Z)$. By the optimality of the coalition-optimal security cost,  $\theta(X) - L(X) + \xi(N\backslash(X\cup Y\cup Z),X) - \xi(X, Y \cup Z) \geq 0$. Now, consider the change in the coalition-optimal security cost $c(S)$ if the set of players in $X$ were to be unsecured. Then, the change in $c(S)$ is given by, $-\theta(X) + L(X) - \xi(N\backslash(X\cup Y\cup Z),X) - \xi(Z, X) + \xi(X,Y)$. Similarly, from the optimality of $c(S)$, $-\theta(X) + L(X) - \xi(N\backslash(X\cup Y\cup Z),X) - \xi(Z, X) + \xi(X,Y) \geq 0$. This implies, $\theta(X) - L(X) + \xi(N\backslash(X\cup Y\cup Z),X) + \xi(X,Y \cup Z) > 0$ from the non-negativity of the security cost parameters. This yields a contradiction, and therefore, $X$ has to be an empty set. Thus, $\Upsilon(S) \subseteq \Upsilon(T)$ and any player $i \in S$ secured under the coalition-optimal security strategy for $S$, i.e., $\Upsilon^1_S = 1$, is also secured under the coalition-optimal security strategy for $T$, i.e., $\Upsilon^1_T = 1$. This completes the proof. \hfill $\square$
\bigskip

\noindent\textbf{Proof of \cref{coopgame_cost}.} Consider $c(S)$, as defined in (\ref{coalitioncost}), and let $\tilde{\sigma}_i$ denote the coalition-optimal security state of player $i$ in coalition $S$. For all $i \in S$ such that $\tilde{\sigma}_i = 1$, $i \in \tilde{S}$. That is, $\tilde{S}$ denotes the set of players in $S$ that are secured under the coalition-optimal security strategy. Further, let us denote by $\tilde{x}_i$, and $\tilde{y}_{ji}$ for $j\in N^-(i)$, the coalition-optimal security actions by player $i \in S$. We note that for all $i \in \tilde{S}$, from (\ref{eqn2}), $\tilde{x}_i = 1$ and $\tilde{y}_{ji} = 1$ for all $j \in N\backslash \tilde{S}$. Further, $y^*_{ji} = 0$ for all $i, j \in \tilde{S}$, since, if players $i$ and $j$ are both secured, it is not optimal (with respect to (\ref{coopgame_cost})) to secure the links between them. Moreover, similarly, for $i \in  N\backslash \tilde{S}$, that is when $i$ is unsecured under the coalition-optimal security strategy, it is not optimal to partially secure $i$ from intrinsic or extrinsic risks. Therefore, for $i \in N\backslash \tilde{S}$, $\tilde{x}_i = 0$ and $\tilde{y}_{ji} = 0$ for all $j \in N\backslash \{i\}$. Thus, 
\begin{align*}
c(S)  &=\ \sum_{i \in S} U_i(\tilde{x}_i, {\boldsymbol{\tilde{y}}_i}| I(i,S))   \nonumber \\
 &=\ 
 \sum_{i \in S} 
 \left( 
    L_i(1-\sigma_i(\tilde{x}_i, {\boldsymbol{\tilde{y}}_i}| I(i,S))) 
    + \theta_i\tilde{x}_i 
    + \sum_{j : (j,i) \in A} \xi_{ji}\tilde{y}_{ji}
 \right) \\
 &=\ \sum_{i \in \tilde{S}} 
 \left( \theta_i + \sum_{j \in N\backslash \tilde{S},\ (j,i) \in A} \xi_{ji} \right) +  \sum_{i \in S\backslash \tilde{S}} L_i.
 \end{align*}
 
 Now, consider the auxiliary network $\graph^*$ and the minimum weight directed cut $(X, \bar{X})$ separating the node $\ell$ and the node set $\{s\} \cup N\backslash S$ in $\graph^*$ with $\{s\} \cup N\backslash S \in X$ and sink $\ell \in \bar{X}$. This constrained minimum weight directed cut in this network identifies $X$ and $\bar{X}$ such that the sum of weights on arcs directed from $X$ to $\bar{X}$ is minimized. The sum of weights of these arcs is given by, 
 \begin{align*}
w(X,\bar{X}) = \sum_{i \in \bar{X}} \left(\theta_i +  \sum_{j \in X, (j,i) \in A} \xi_{ji} \right) +  \sum_{i \in X} L_i.  \nonumber 
 \end{align*}
From comparing the expressions, $w(X,\bar{X})$ and $c(S)$ are simultaneously minimized when $\bar{X} = \tilde{S} \cup \{\ell\}$ and $X = N\backslash \tilde{S}$. This completes the proof. 
\hfill $\square$
\bigskip

\noindent\textbf{Proof of \cref{convexity}.} Consider coalitions $S$ and $T$ such that $T = S \cup \{j\}$ and $i \notin S$. Denote $S' = S\cup \{i\}$ and $T' = T\cup\{i\}$. Suppose that $i \notin \Upsilon(T')$, that is player $i$ is not secured in the coalition $T'$, then, from \cref{monotone}, player $i$ is not secured in the coalition $S'$ either, $i \notin \Upsilon(S')$. Therefore, $c(T\cup\{i\}) = c(T) + L_i$ and $c(S\cup\{i\}) = c(S) + L_i$. Thus, $c(T\cup\{i\}) - c(T) = c(S\cup\{i\}) - c(S)$. Suppose instead that $i \in \Upsilon(T')$ but $i \notin \Upsilon(S')$, that is, $i$ is secured in the coalition $T'$ whereas it is unsecured in the coalition $S'$. Then, $c(S\cup\{i\}) = c(S) + L_i$. Moreover, from (\ref{coalitioncost}), $c(T\cup\{i\}) \leq c(T) + L_i$. Therefore, $c(S\cup\{i\}) - c(S) \geq c(T\cup\{i\}) - c(T)$. 

\begin{figure}[!htb]
  \begin{center}
    \includegraphics[scale = 0.20]{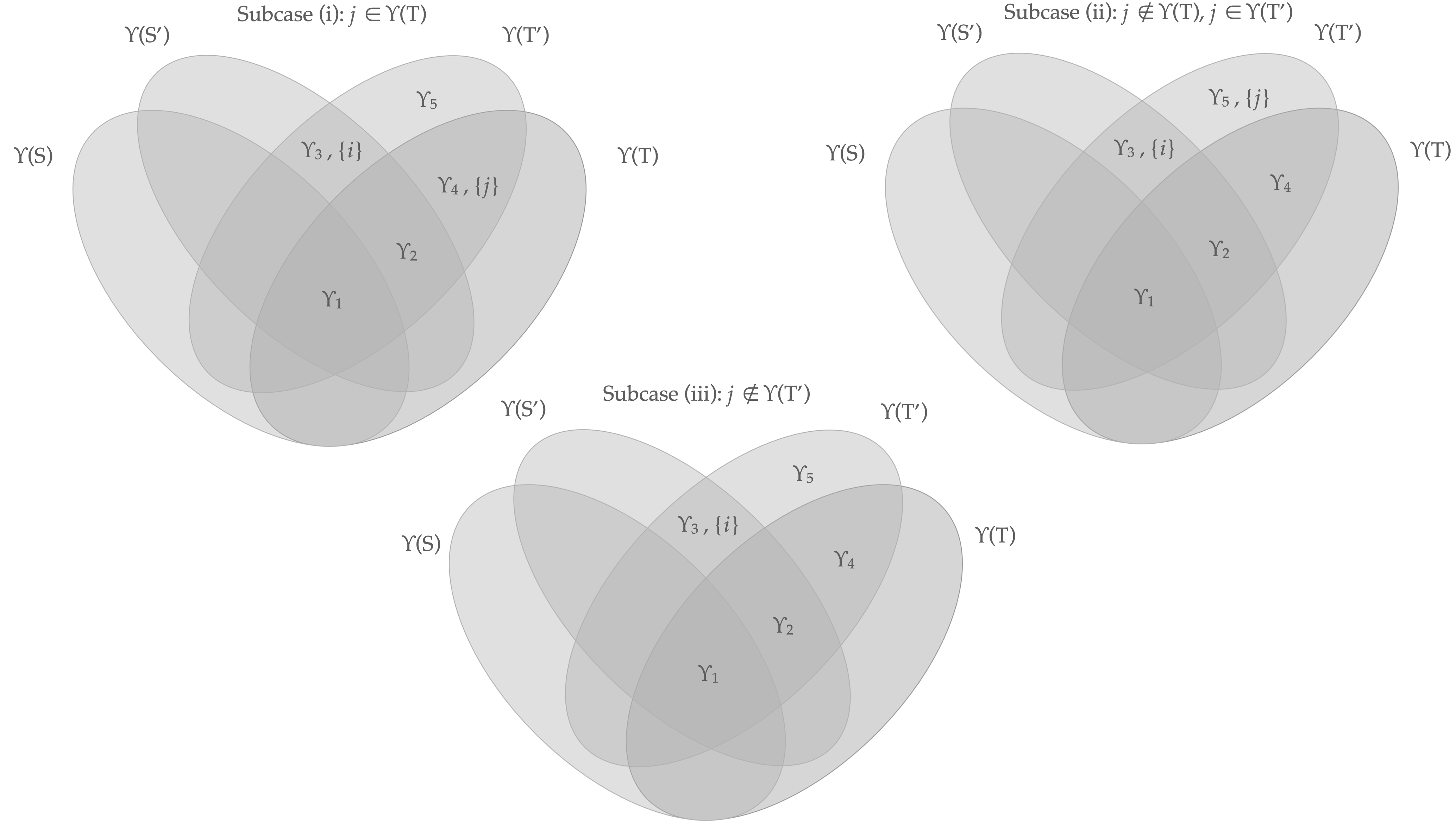}
    \caption{Intersections of the secured sets of coalitions $S$, $S'$, $T$, and $T'$.}
    \label{fig_convex}
  \end{center}
\end{figure}

Finally, suppose $i \in \Upsilon(S')$. Then, again, from \cref{monotone}, $i \in \Upsilon(T')$. Let us denote $\Upsilon(S) = \Upsilon_1$. We then consider the intersections of the secured sets of the coalitions $S'$, $T$, and $T'$ as follows. Denote $\Upsilon(S')\cap \Upsilon(T) \cap \Upsilon(T') = \Upsilon_2$, $\left(\Upsilon(S')\cap \Upsilon(T') \cap \left(N\backslash \Upsilon(T)\right)\right) \backslash \{i\} = \Upsilon_3$, $\Upsilon(T)\cap \Upsilon(T') \cap \left(N\backslash \Upsilon(S')\right)  \backslash \{j\} = \Upsilon_4$, and $\Upsilon(T')\cap \left(N\backslash \Upsilon(T)\right) \cap \left(N\backslash \Upsilon(S')\right) \backslash \{i, j\} = \Upsilon_5$. Furthermore, let $\Upsilon_6 = T' \backslash \Upsilon(T')$ and $\Upsilon_7 = N\backslash T'$. Note that, by construction, $\Upsilon_1$, $\Upsilon_2$, $\Upsilon_3$, $\Upsilon_4$, $\Upsilon_5$, $\Upsilon_6$, $\Upsilon_7$ are all disjoint. The intersections of $\Upsilon(S)$, $\Upsilon(S)$, $\Upsilon(S)$, and $\Upsilon(S)$ are depicted for clarity in \cref{fig_convex}. We now consider three exhaustive subcases. 

In subcase (i), we consider $j \in \Upsilon(T)$. This implies $j \in \Upsilon(T')$ as well. Further, since, by \cref{monotone}, all players secured in $S'$ are also secured in $T'$, it therefore follows that $\Upsilon(S') = \Upsilon_1 \cup \Upsilon_2 \cup \Upsilon_3 \cup \{i\}$. Similarly, players secured in $T$ are also secured in $T'$. Therefore, $\Upsilon(T) = \Upsilon_1 \cup \Upsilon_2 \cup \Upsilon_4 \cup \{j\}$. Finally, by construction, $\Upsilon(T') = \Upsilon_1 \cup \Upsilon_2 \cup \Upsilon_3 \cup \Upsilon_4 \cup \Upsilon_5 \cup \{i, j\}$. For sets of players $X$ and $Y$ in $N$, denote $\theta(X) = \Sigma_{k \in X}\ \theta_k$ and $\xi(X,Y) = \Sigma_{k \in X, l \in Y}\ \xi_{kl}$. Then,
$c(T') - c(T) = \theta(\Upsilon_3 \cup \Upsilon_5 \cup \{i\}) - L(\Upsilon_3 \cup \Upsilon_5) + \xi(\Upsilon_6, \Upsilon_3 \cup \Upsilon_5 \cup \{i\}) + \xi(\Upsilon_7, \Upsilon_3 \cup \Upsilon_5 \cup \{i\}) - \xi(\Upsilon_3 \cup \Upsilon_5, \Upsilon_1 \cup \Upsilon_2 \cup \Upsilon_4 \cup \{j\}) - \xi(\{i\}, \Upsilon_1 \cup \Upsilon_2 \cup \Upsilon_4 \cup \{j\})$. Further, $c(S') - c(S) = \theta(\Upsilon_2 \cup \Upsilon_3 \cup \{i\}) - L(\Upsilon_2 \cup \Upsilon_3) + \xi(\Upsilon_4\cup\Upsilon_5\cup\Upsilon_6,\Upsilon_2\cup\Upsilon_3\cup\{i\}) + \xi(\Upsilon_7 \cup \{j\}, \Upsilon_2 \cup \Upsilon_3 \cup \{i\}) - \xi(\Upsilon_2\cup\Upsilon_3\cup\{i\},\Upsilon_1)$. Let us suppose that $c(S') - c(S) \leq c(T') - c(T)$. Then, from simplifying the above two expressions, we obtain: 
\begin{multline}
\theta(\Upsilon_2) - L(\Upsilon_2) + \xi(\Upsilon_4\cup\Upsilon_5\cup\Upsilon_6,\Upsilon_2) + \xi(\Upsilon_4\cup\Upsilon_5,\Upsilon_2\cup\Upsilon_3\cup\{i\}) + \xi(\Upsilon_7,\Upsilon_2) + \xi(\{j\},\Upsilon_2\cup\Upsilon_3\cup\{i\}) \\ -\xi(\Upsilon_2,\Upsilon_1) -\theta(\Upsilon_5)+L(\Upsilon_5)-\xi(\Upsilon_6,\Upsilon_5)-\xi(\Upsilon_7,\Upsilon_5)+ \xi(\Upsilon_3,\Upsilon_2\cup\Upsilon_4\cup\{j\})+ \\ \xi(\Upsilon_3,\Upsilon_2\cup\Upsilon_4\cup\{j\})+\xi(\Upsilon_5,\Upsilon_1\cup\Upsilon_2\cup\Upsilon_4\cup\{j\})+\xi(\{i\},\Upsilon_2\cup\Upsilon_4\cup\{j\}) \leq 0. 
\end{multline}

Now, from the optimality of $\Upsilon(S) = \Upsilon_1$, we obtain that if the set of players in $ \Upsilon_2$ were also to be secured in $S$, then, 
\begin{equation}
\theta(\Upsilon_2) - L(\Upsilon_2) + \xi(\Upsilon_3\cup\Upsilon_4\cup\Upsilon_5\cup\Upsilon_6,\Upsilon_2) - \xi(\Upsilon_2,\Upsilon_1)+\xi(\Upsilon_7\cup\{i\}\cup\{j\},\Upsilon_2) \geq 0.
\end{equation}
Similarly, from the optimality of $\Upsilon(T') = \Upsilon_1 \cup \Upsilon_2 \cup \Upsilon_3 \cup \Upsilon_4 \cup \Upsilon_5 \cup \{i, j\}$, if the set of players in $\Upsilon_5$ were to be unsecured instead, then, 
\begin{equation}
L(\Upsilon_5)-\theta(\Upsilon_5) - \xi(\Upsilon_6,\Upsilon_5) - \xi(\Upsilon_7,\Upsilon_5) + \xi(\Upsilon_5,\Upsilon_1\cup\Upsilon_2\cup\Upsilon_4\cup\{i\}\cup\{j\}) \geq 0.
\end{equation}
Summing (8)-(10), we obtain, 
\begin{equation}
\xi(\Upsilon_4, \Upsilon_2\cup\Upsilon_4\cup\{i\}) + \xi(\Upsilon_5\cup\{j\},\Upsilon_2\cup\Upsilon_3) + \xi(\Upsilon_3\cup\{i\},\Upsilon_4\cup\{j\}) \leq 0,
\end{equation}
which yields a contradiction since by assumption, parameters $\theta_i$ and $\xi_{ji}$ are positive for all $i \in N$ and $(j,i) \in A$. Thus, $c(S') - c(S) \geq c(T')-c(T)$.  

In subcase (ii), we consider $j \notin \Upsilon(T)$ but $j \in \Upsilon(T')$ and in subcase (iii), we consider $j \notin \Upsilon(T')$. Using arguments similar to subcase (i), we can demonstrate that in both these subcases, $c(T')-c(T) \leq c(S')-c(S)$. Furthermore, by induction, for any coalitions $S$ and $T$ such that $S \subset T \subset N$, and for $i \notin T$, $c(T \cup \{i\}) - c(T) \leq c(S\cup \{i\}) - c(S)$. Thus, the coalition-optimal security cost $c(S)$ is submodular in $S$. 

Finally, therefore it follows from Shapley (1971), that the interdependent security cost sharing game has a non-empty core, i.e., there exists a stable security cost sharing mechanism.
\hfill $\square$
\bigskip

The following observation notes that if a player is unsecured under the network-optimal security strategy, then, the player is allocated $L_i$ by all stable cost sharing arrangements.  

\begin{lemma}\label{restrict_assump}
Consider $i \in N$ such that $i \notin \sStar$ and an arbitrary core allocation $\phi$ of the interdependent security cost sharing game on network $\graph$.
\begin{enumerate}
    \item[i.] $\phi$ allocates to player $i$, $\phi_i = L_i$.
    \item[ii.] Define $\graph'$ as the induced subgraph of $\graph$ on the node set $N\backslash \{i\}$. Further, let $\theta'_j = \theta_j + \xi_{ij}$ for $j \in N^+(i)$, and let all the other security cost parameters of $\graph'$ be identical to the corresponding costs in $\graph$. Then, there exists a one-to-one correspondence between the core allocations of the interdependent security games on $\graph$ and $\graph'$, respectively.
    \end{enumerate}
\end{lemma}

While analyzing security cost sharing mechanisms, \cref{restrict_assump} also allows us to restrict our attention to networks $\mathbb{G}$ and associated cost parameter vectors such that all firms are secured under the network-optimal security strategy.  
\smallskip 

\noindent\textbf{Proof of \cref{restrict_assump}.} Consider $i \in N$ such that $i \notin \sStar$. That is, $i$ is not secured under the network-optimal security strategy. Then, suppose $\phi$ is an arbitrary core allocation of the corresponding interdependent security cost sharing game. Suppose the cost allocated to $i$ by $\phi$, $\phi_i > L_i$. This leads to a contradiction since $c(\{i\}) = L_i$ by \cref{prop_network} and $c(\{i\}) < \phi_i$ implying $\phi$ cannot be a core allocation. Suppose instead that $\phi_i < L_i$. Note that, since $i \notin \sStar$, $c(N\backslash\{i\}) = c(N) - L_i$. Since $\phi$ belongs to the core, it is an efficient allocation, and therefore, $\phi(N\backslash \{i\}) = c(N) - \phi_i > c(N) - L_i = c(N\backslash\{i\})$, again leading to a contradiction to the coalitional rationality of core allocations. Thus, $\phi_i = L_i$. This completes the proof of part (i) of the lemma. 

Consider the associated interdependent security cost sharing game $(N\backslash\{i\}, c')$ defined on $\graph'$. Since $i \notin \sStar$, it follows from \cref{monotone} that it is not secured under the coalition-optimal security strategy for any coalition in $N$. Thus, it follows from (1) that for any player $j$ that is secured under a coalition-optimal security strategy for any coalition, $y_{ij} = 1$. Therefore, from (4), for any subset $S\subseteq N\backslash\{i\}$, it follows that $c'(S) = c(S)$. Finally, from part (i) of the lemma, since player $i$ is allocated $L_i$ by all core allocations in $(N,c)$, for any core allocation $\phi'$ in $(N\backslash\{i\}, c')$, consider its extension to an allocation $\phi$ in $(N,c)$ such that $\phi_j = \phi'_j$ for all $j \in N\backslash \{i\}$ and $\phi_i = L_i$. It follows that since $\phi'$ is a core allocation in $(N\backslash\{i\}, c')$, $\phi$ is efficient and also satisfies all the core inequalities given by (7) in $(N,c)$. The reverse direction also follows by identical arguments. This demonstrates a one-to-one correspondence between the core allocations of $(N,c)$ and $(N\backslash\{i\}, c')$. \hfill $\square$

\bigskip

\noindent\textbf{Proof of \cref{shapley_np}.} Consider an assembly network (or, also termed a star network), where $N$ denotes the set of players in the network, and the set of arcs $A = \{ (i,j): i \in N\backslash\{0\}, j = \{0\})$. Thus, node $0$ corresponds to the assembler in the network (or, the central node in the star network). Further, we assume that $L_i$ for all $i \neq 0$ is sufficiently large such that it is optimal for $i$ to be independently secured. Further, assume that $\theta_0 + \Sigma_{i \in N\backslash\{0\}} \xi_{i0} > L_0$ and therefore, node $0$ will not be secured independently. Also, assume that $L_0 > \theta_0$. Since all the other nodes in $N$ will be secured under the network-optimal security strategy, it is also optimal for $0$ to be secured under the network-optimal security strategy. 
Also, therefore, note that player $0$ will be secured in a coalition $S$ that contains $0$, i.e., $0 \in \Upsilon(S)$ if and only if $\theta_0 + \Sigma_{i \in N\backslash S}\ \xi_{i0} < L_0$.

Consider $i \neq 0$ in $N$. Then, 
\begin{equation}
c(S\cup \{i\}) - c(S) = 
\begin{cases}
    \theta_i, &\text{ if } 0 \notin S,\\
    \theta_i, &\text{ if } 0 \in S, 0 \notin \Upsilon(S\cup\{i\}),\\
        \theta_i - \xi_{i0}, &\text{ if } 0 \notin \Upsilon(S), 0 \in \Upsilon(S\cup\{i\}).
  \end{cases}
  \label{eqn_a1}
\end{equation}
If $\Phi_i$ denotes the security cost allocated to player $i$ by the Shapley value based security cost sharing mechanism, then, from (12) and (16), $\Phi_i < \theta_i$ if and only if there exists a subset $T$ in $N\backslash \{i\}$, where $T = N\backslash (S\cup \{i\})$, such that $\Sigma_{i \in T}\ \xi_{i0} \in [L_0 - \xi_{i0} - \theta_0, L_0 - \theta_0)$. This is a generalization of the classical subset sum problem in which given a set of integers, the problem is to identify whether there exists a subset that sums to a pre-specified target value. Since, the classical subset sum problem is well-known to be NP-complete, it follows that determining whether $\Phi_i < \theta_i$ is also NP-complete. This concludes the proof. \hfill $\square$
\bigskip

\noindent\textbf{Proof of \cref{shapley_special}.} Note that when $L_i > \theta_i + \Sigma_{j \in N^-(i)} \xi_{ji}$, then, from \cref{prop_independent} and \cref{monotone}, player $i \in S$ is always secured under the coalition-optimal security strategy for any such coalition $S$. Therefore, it follows from (4) that $c(S) = \Sigma_{i \in S} \left(\theta_i + \Sigma_{j \in N^-(i), j \in N\backslash S}\  \xi_{ji}\right)$. Now, for each $i \in N$, define a corresponding cooperative game given by the characteristic function $c^i(S)$ for $S \subseteq N$ as follows:
\begin{equation}
c^i(S) =  
\begin{cases}
    \theta_i + \sum\limits_{j \in N^-(i)} \xi_{ji}, &\text{ if } i \in S,\\
     0, &\text{ if } i \notin S.
  \end{cases}
  \label{eqn_a2}
\end{equation}
It can be easily seen that the Shapley value of $(N,c^i)$ allocates to player $i$, $\Phi_i(N,c^i) = \theta_i + \sum\limits_{j \in N^-(i)} \xi_{ji}$ and for all players $j \neq i$, $\Phi_j(N,c^i) = 0$. Furthermore, for each arc $a = (j,i) \in A$, define a corresponding cooperative game given by the characteristic function $c^a(S)$ for $S\subseteq N$ as follows: 
\begin{equation}
c^a(S) =  
\begin{cases}
    - \xi_{ji}, &\text{ if } i, j \in S,\\
     0, &\text{ otherwise. } 
  \end{cases}
  \label{eqn_a3}
\end{equation}
Again, from the symmetry property, it can be seen that the Shapley value of $(N,c^a)$ allocates to players $i$ and $j$, $\Phi_i(N,c^a) = \Phi_j(N,c^a) = - \xi_{ji}/2$ and for all players $k \neq i, j$, $\Phi_k(N,c^i) = 0$. Finally, note that for each $S \subseteq N$, $c(S) = \Sigma_{i\in N}\ c^i(S) + \Sigma_{a \in A}\ c^a(S)$. Therefore, from the additivity property, the Shapley value based security cost allocation is given by, \begin{equation*} \Phi_i = \Phi_i(N,c^i) + \sum\limits_{a \in A} \Phi_i(N,c^a) = \theta_i + \sum_{j \in N^-(i)} \frac{\xi_{ji}}{2} - \sum_{j \in N^+(i)} \frac{\xi_{ij}}{2}. \end{equation*} \hfill $\square$
\bigskip

\noindent\textbf{Proof of \cref{extreme_core}.} \cref{convexity} demonstrates that the interdependent security cost sharing game is convex. Further, from Theorems 3 and 5 in \citet{shapley1971}, the allocation $x_\pi$, also sometimes denoted as marginal worth vectors, is an extreme point of the core. Finally, for any permutation $\pi$ of $N$, from \cref{coopgame_cost}, both $c(\{\pi_1,\pi_2,...,\pi_i\})$ and $c(\{\pi_1,\pi_2,...,\pi_{i-1}\})$ can be computed in polynomial time, and therefore, so can $x_{\pi_i}$ for all $i \in N$. \hfill $\square$
\bigskip

\noindent\textbf{Proof of \cref{coresymmetry}.} We demonstrate that the security cost sharing allocation $x_\pi$ does not, in general, satisfy the symmetry property, by considering a 2-player example. Consider a network given by the node set $N = \{1, 2\}$ and the arc set $S = \{(1,2), (2,1)\}$. Further, let the security cost parameters be given by $\theta_1 = \theta_2 = 5$, and $\xi_{12} = \xi_{21} = 10$, and finally, $L_1 = L_2 = 100$. Clearly, the security incurred by each coalition, i.e., the characteristic cost function of the associated cooperative game is given by $c(\phi) = 0$, $c(\{1\}) = c(\{2\}) = 15$, and $c(\{1,2\}) = 10$. Both the players $1$ and $2$ in the game satisfy the condition that for each subset $S \subset N$, $c(S\cup \{1\}) - c(S) = c(S\cup \{2\}) - c(S)$. However, consider the permutation of the players given by $\pi = \{1, 2\}$. Then, $x_{\pi_1} = c(\{1\}) - c(\phi) = 15$, whereas, $x_{\pi_2} = c(\{1,2\}) - c(1) = -5$. This example demonstrates that the security cost-sharing allocation $x_\pi$ does not necessarily satisfy the symmetry property. \hfill $\square$
\bigskip

\noindent\textbf{Proof of \cref{shapley_bilateral}.} Denote $\Delta(S, i) := c(S\cup \{i\}) - c(S)$. Indeed, the Shapley value allocates to player $i$, 
\begin{equation}
\Phi_{i} 
= \sum_{S\subseteq N\setminus\{i\}} k(|S|, n)\Delta(S,i),
\end{equation} 

where $k(|S|, n) = \frac{|S|!(n-|S|-1)!}{n!}$. Note that $k(|S|,n)$ depends only on $|S|$ and not on the security cost parameters of the players in $S$.

\textbf{Part (i).} We consider the computation of $\Delta(S, i)$. We prove that if the given conditions hold, then the Shapley value is bilaterally implementable. Suppose $i \notin  \bar \minSecG(j)$ for all $j \in N(i)$. Then, for all $S$, the security states of players in $S$ remains the same under the coalitionally optimal strategy for $S \cup \{i\}$. Therefore, $\Delta(S, i)$ is either $L_i$ or $\theta_i + \sum_{j \in S\backslash \Upsilon(S), j \in N^-(i)} \xi_{ji} - \sum_{j \in \Upsilon(S), j \in N^+(i)} \xi_{ij}$. In both cases, $\Delta(S,i)$ only contains parameters involving $i$ and its direct partners. 

Then, suppose $i \in \bar \minSecG(j)$ for some neighbors $j$, but $|N(j)|  = 1$ for all such $j$, that is, $i$ is the only direct partner of $j$ whenever it belongs to a minimal coalitionally rational set for $j$. Then, again, by similar arguments as before, $\Delta(S,i)$ is either $-L_i$ or only contains parameters involving $i$ and other direct partners of $i$. Therefore, it follows, from (20), that $\Phi$ is bilaterally implementable. 

\textbf{Part (ii).} Now, suppose there exists a player $i \in N$ such that $i \in \bar \minSecG(j)$ for some $j \in N(i)$ such that $|N(j)\backslash N(i)| > 1$. That is, suppose there exists some $P$ such that $P\cup \{i\}$ is a minimal coalitionally rational security set for $j$. Then, clearly, there exist sets $S$ such that $P \subset S$, $j \in S$ and a neighbor of $j$ that is not a neighbor of $i$, say $k$, that is, $k \in N(j)\backslash N(i)$ also belongs to $S$. Also, suppose $i \notin S$ and $S$ does not contain any other minimal coalitionally rational security set for $j$. Then, $j \notin \Upsilon(S)$ but $j \in \Upsilon(S\cup\{i\})$. Now, we have three possible cases. If $k \in \Upsilon(S)$, then, $\Delta(S, i)$  will contain the term $-\xi_{jk}$. If $k \notin \Upsilon(S)$ but $k \in \Upsilon(S\cup\{i\})$, then, $\Delta(S, i)$ will contain the terms $L_k$ and $\theta_k$ where $k \notin N(i)$. Finally, if $k\notin \Upsilon(S\cup\{i\})$, then,  $\Delta(S, i)$ will include the term $\theta_{kj}$. Thus, in all three cases, from (20), it also follows that $\Phi_i$ will include linear terms involving a player $k$ that is not a partner of $i$, thereby violating the bilateral implementability of $\Phi$. 
\hfill $\square$
\bigskip

\noindent\textbf{Proof of \cref{prop:BilateralSymmAlloc}.} The agreeable allocation, denoted by $x^\star$, belongs to the core since it is a convex combination of a subset of extreme points of the core. We now demonstrate that it also satisfies all the other properties.
\smallskip

\paragraph{i. Polynomial-time computability. }
We first present an algorithm that computes $\coreSubset_1$ to $\coreSubset_\ell$ in polynomial time. Then, given $\coreSubset_1$ to $\coreSubset_\ell$, we provide a closed-form expression for the allocation $x^\star$. Denote $n = |N|$. From \cref{prop_independent}, it follows that the set $\coreSubset_1$ can be identified in polynomial time by checking whether $L_i \geq \theta_i + \Sigma_{j \in N^-(i)} \xi_{ji}$ for each $i \in N$. If $\coreSubset_1 = \emptyset$, then no agreeable permutation exists. Iteratively, suppose for $k \ge 2$, $\coreSubset_1,\dots,\coreSubset_{k-1}$ is known. Compute $\bar{\coreSubset_{k-1}}$. If $\bar{\coreSubset_{k-1}} = N$, then we terminate with $\ell = k-1$. Otherwise, for each $i\in N\setminus\bar{\coreSubset_{k-1}}$, if $\security[\bar{\coreSubset_{k-1}}\cup\{i\}][i] =1 $, then, $i \in \coreSubset_k$. To verify whether $\security[\bar{\coreSubset_{k-1}}\cup\{i\}][i] =1$, since all players in $\coreSubset_{k-1}$ are secured, it involves comparing $L_i$ with $\theta_i + \Sigma_{j \in N\backslash\bar{\coreSubset_{k-1}} \cap N^-(i)}\ \xi_{ji} - \Sigma_{j \in \bar{\coreSubset_{k-1}}\cap N^+(i)}\ \xi_{ij}$. Thus, $\coreSubset_k$ can be constructed in polynomial time. If $\coreSubset_k = \emptyset$, then again, no agreeable permutation exists. 

Now that we have $\coreSubset_k$ for $k=1,\dots,\ell$, we note that the extreme core allocation corresponding to any agreeable permutation allocates to a player $i \in \coreSubset_k$, (i) its own cost of intrinsic security, $\theta_i$, (ii) the cost of extrinsic security with respect to players not in $\bar{\coreSubset_k}$, (iii) the security cost savings generated for the players in $\bar{\coreSubset_{k-1}}$, (iv) finally, the cost of extrinsic security and the security cost savings generated with respect to its partners also in $\coreSubset_k$. Consider partners $i$ and $j$ in $\coreSubset_k$. For exactly half of the agreeable permutations, $i$ shall appear before $j$ in the permutation, whereas for exactly half the permutations, $j$ shall appear before $i$. Since the agreeable allocation is a convex combination of the extreme core allocations induced by all agreeable permutations, we have,

\begin{align}
  x^\star_i =\ \ \theta_i +  \sum_{\substack{j\in N^-(i)\\ j \in (N\setminus \bar{\coreSubset_k})}} \xi_{ji} - \sum_{\substack{j\in N^+(i) \\ j \in \bar{\coreSubset_{k-1}}}} \xi_{ij} + \sum_{\substack{j\in N^-(i) \\ j \in \coreSubset_k}} \frac{\xi_{ji}}{2} - \sum_{\substack{j\in N^+(i) \\ j \in \coreSubset_k}} \frac{\xi_{ij}}{2}   & \text { if }i \in \coreSubset_k. \label{eq:BilateralSymmAlloc}
\end{align}
\smallskip

\paragraph{ii. Efficiency. }Efficiency follows from the fact that the defined allocation is a convex combination of efficient allocations. 
\smallskip

\paragraph{iii. Marginality property. } Observe that in the definition of the allocation defined by a permutation, the allocation is always the marginal contribution of the player $i$ to the coalition of every player that appears earlier in the permutation. Clearly, this has the marginality property. Since $x^\star$ is a linear combination of such marginal allocation, $x^\star$ has the marginality property too.
\smallskip 

\paragraph{iv. Null player property. } Observe that for every allocation defined by some permutation $\pi$, the payoff of a player $i$ is $c(S\cup\{i\}) - c(S)$ for some $S$. This difference is a constant $\bar c$ if $i$ is a dummy player. The proposed allocation suggests a convex combination of these payoffs, which, in this cases is a convex combination $\bar c$ which is $\bar c$.
\smallskip 

\paragraph{v. Symmetry. }Observe that if two players $i$ and $j$ are symmetric, then $\exists\, k$ such that $\{i,j\} \subseteq \coreSubset_{k}$. But, now, from (\ref{eq:BilateralSymmAlloc}), it follows that their security cost allocations are identical.
\smallskip 

\paragraph{vi. Bilateral Implementability. } Bilateral implementability, again, follows directly from inspection of (\ref{eq:BilateralSymmAlloc}). \hfill$\square$
\bigskip 

\newpage

\section{Multilateral Implementability and $\delta$-Agreeable Allocations}
\medskip 

In this section, our first objective is to identify a subset of permutations of the players in $N$ that we denote as $\delta$-agreeable permutations. We now introduce some preliminary definitions and notions that aid us in constructing $\delta$-agreeable permutations. We again limit our attention to networks where all firms are secured in the grand coalition. 

Consider the network $\graph = (N,A)$ with, as before, $\theta_i$ and $L_i$ representing the intrinsic security cost and expected penalty in case of a realized risk for each $i\in N$, and $\xi_{ji}$ denoting the cost of extrinsic security for the arc $(j,i)\in A$. Further, let $\delta \geq 1$ be an integer. Then, for a  given subset $S\subseteq N$, we define a {\it $\delta$-rational security set}, $R_\delta(S)$, as follows. 
  \begin{align}\label{eqn_EC10}
    R_\delta \left( S \right) \quad&:=\quad \left\{ 
      V \subseteq (N\setminus  S) : c( S \cup V ) < c( S) + \sum_{v\in V} L_v, |V| = \delta
    \right\}
  \end{align}

In other words, $R_\delta(S)$ consists of all $\delta$-sized subsets such that if all members of a subset join the coalition $S$ of players, then, the total security cost of the resulting coalition will be strictly smaller than if they were not a part of the coalition. Indeed, this can happen only if at least one of the $\delta$ elements  is secured as a result of joining the coalition with $S$.

We next define {\it $\delta$-minimal rational security sets} ($\delta$-MRS) for a coalition $S$, $\mathcal{R}_\delta(S)$, as the $\delta^*$-rational security sets, $R_{\delta^*}(S)$ for $\delta^* \leq \delta$ such that $R_{\delta^'}(S)$ is empty for all $\delta^' < \delta^*$ and $R_{\delta^*}(S)$ is non-empty. In this case, we say that the $\delta$-MRS is achieved for $\delta = \delta^*$. Further, we drop the reference to $S$, if it is clear from the context.

\begin{lemma} \label{lem:allSecMinRat}
  Given the network $\graph$ and a coalition $S$, for each $\delta^*$-sized subset $\{v_1, \dots, v_{\delta^*}\} \in \mathcal{R}_\delta(S)$, each of the players in $v_1,\dots, v_{\delta^*}$ is secured in the coalition $S \cup \{v_1, \dots, v_{\delta^*}\}$.
\end{lemma}
\textbf{Proof of \cref{lem:allSecMinRat}.} Suppose there exists $v_i$ such that $v_i$ is not secured in the coalition $S \cup \{v_1, \dots, v_{\delta^*}\}$. Then, $c(S \cup \{v_1, \dots, v_{\delta^*}\}) = c(S \cup \{v_1, \dots, v_{\delta^*}\} \setminus \{v_i\}) + L_{v_i}$. Then, from (\ref{eqn_EC10}), it follows that the set $\{v_1, \dots, v_{\delta^*}\} \setminus \{v_i\}$ is also a $\delta$-rational security set contradicting the minimality of $\{v_1, \dots, v_{\delta^*}\}$.
\hfill $\square$
\medskip

\begin{lemma} \label{lem:MRSconnected}
  The players in any $\delta$-minimal rational security set of a coalition $S$ constitute a connected set of nodes in $\graph$. 
\end{lemma}
\textbf{Proof of \cref{lem:MRSconnected}.}
  Suppose that a $\delta$-minimal rational security set, $B$, does not correspond to a connected set of nodes in $\graph$. That is, then, $B = B_1 \cup B_2$, where $B_1$ is a connected set of nodes, neither $B_1$ nor $B_2$ is empty, such that there is no arc from $B_1$ to $B_2$ or vice versa. From \cref{lem:allSecMinRat}, it follows that every node in $B$ is secured. However, since there are no arcs between the players in $B_1$ and $B_2$, it follows that in the coalition $S \cup B_1$, all players in $B_1$ will  be secured implying that $B_1$ is already a $\delta'$-rational set with $\delta' < |B|$, contradicting the minimality of $B$.
\hfill $\square$

We now consider the $\delta$-minimal rational security sets ($\delta$-MRS) for a given coalition $S$ and describe a procedure in \cref{alg:validPerm} that allows us to augment the coalition $S$ with a specific set of permutations of the players in the $\delta$-MRS. This augmenting procedure will then in turn be used in constructing $\delta$-agreeable permutations in \cref{alg:rhoAgree}.

Suppose $\mathcal{R}_\delta(S) = \{B_1, B_2, \dots, B_k\}$ where each $B_i\subseteq N$ can potentially contain common elements, i.e., there may exist players belonging to several $\delta$-minimal rational security sets for a given coalition $S$. For a set $T$, let $\pi(T)$ denote the set of permutations of the elements in the set $T$ and for $\pi \in \pi(T)$, let $\pi(i)$ denote the $i^{th}$ element in the permutation $\pi$.  

\begin{algorithm}[h]
\begin{algorithmic}
\Ensure{A valid permutation $\zeta$ of players that appear in at least one of the sets in $\mathcal{R}_\delta(S)$}\\
$\zeta \gets \emptyset$\;\\
$\pi \in \pi(\{1,\cdots,k\})$\;
\For{$i = 1,\cdots,k$}\;
 \State   $C_i := \{v\in B_{\pi(i)}: v \notin B_{\pi(1)} \cup \dots \cup B_{\pi(i-1)} \} $\;
 \State     $\pi' \in \pi(C_i)$
  \State    Append $\pi'$ to $\zeta$\;
\EndFor
\end{algorithmic}
\caption{Augmenting valid permutations corresponding to $\delta$-MRS} \label{alg:validPerm}
\end{algorithm}

\begin{lemma} \label{lem:validPermSecured}
Consider a valid permutation $\zeta$ of players that appear in at least one of the sets in $\mathcal{R}_\delta(S)$ obtained from \cref{alg:validPerm}. Suppose $\zeta = (v^1_1, v^1_2, v^1_3,\dots,v^1_{\ell_1}, v^2_1,\dots,v^2_{\ell_2},\dots,v^k_1,\dots,v^k_{\ell_k})$ where $v^i_j \in C_i$ for all $j$ and where $C_i$ is as defined in \Cref{alg:validPerm}. 
  Then, for each $i=1,\dots,k$, \emph{all} players are secured in the coalition $S \cup \{ 
  v^1_1,\dots, v^i_{\ell_i} \}$.
\end{lemma}
\noindent\textbf{Proof of \cref{lem:validPermSecured}.}
  From \Cref{lem:allSecMinRat}, we know that for any $j$, all nodes in $S \cup \{v^j_1,\dots,v^j_{\ell_j}\}$ are secured. 
  Observe that any coalition of the form $S \cup\{v^1_1,\dots, v^i_{\ell_i} \} $ is  the union of $S$ and $i$ $\delta$-MRS sets of $S$. When players join a coalition, we know, from \cref{monotone} that players that are secured in the original coalition continue to remain secured. This implies all players in $S \cup\{v^1_1,\dots, v^i_{\ell_i} \} $ are secured. This completes the proof. 
\hfill $\square$
\medskip

We are now in a position to employ the notion of valid permutations to construct a $\delta$-agreeable permutation. For clarity, let us denote a valid permutation of players that appear in at least one of the sets in $\mathcal{R}_\delta(S)$ obtained from \cref{alg:validPerm} by $\zeta(S)$.
\newpage

\begin{algorithm}[h]
\begin{algorithmic}
\Ensure{A $\delta$-agreeable permutation of the players in $\graph$}\\
$S_0 \gets \emptyset$\;\\
$i \gets 0$\;\\
$\phi \gets \emptyset$\;
\While{$S_i \neq N$}\;
  \If {$\mathcal{R}_\delta(S_i) = \emptyset$}
  \State $\delta$-agreeable permutations do not exist. 
  \EndIf
  \State  Append $\zeta(S_i)$ to $\phi$\;
   \State   $i \gets i + 1 $\;
  \State  $S_i \gets S_{i-1} \cup \zeta(S_{i-1})$\;
  \EndWhile
\end{algorithmic}
\caption{Constructing a $\delta$-agreeable permutation}\label{alg:rhoAgree}
\end{algorithm}

\begin{proposition}\label{prop:ec1}
The extreme core allocation corresponding to any $\delta$-agreeable permutation is (i) efficient, (ii) belongs to the core, (iii) is  $(\delta+1)$-laterally implementable, (iv) is  polynomial-time computable in $|N|$ (could be exponential in $\delta$), (v) satisfies  marginality and (vi) satisfies the null player property. 
\end{proposition}

\noindent \textbf{Proof of \cref{prop:ec1}.}
  (i), (ii), (v), and (vi) follow immediately given that the allocation is an extreme core allocation based on permuting the set of players in a convex cooperative game. (iv) follows since given any $S$, one can find the set of all $\delta$-MRS in time bounded by a polynomial in $N$ since there are at most ${|N|\choose \delta} \sim O(|N|^{\delta+1})$ subsets to check. Whether any given subset is in $\delta$-MRS can be checked in polynomial time since it only involves computing optimal security costs of coalitions.  Thus, the extreme core allocation is poly-time computable. The rest of the proof is dedicated to proving that the allocation is $(\delta+1)$-laterally implementable.

  Let $\varphi$ be a $\delta$-agreeable permutation obtained from \cref{alg:rhoAgree}. 
  From \cref{lem:validPermSecured}, we know that when all players belonging to a particular $C_i$ are added to a coalition, they are all secured. However, since the $C_i$'s are all subsets of $\delta$-MRS, each $C_i$ has at most $\delta$ elements.  This means, any player can remain unsecured until at most $\delta-1$ more players are added to the coalition. However, from \cref{lem:MRSconnected}, we know that each of $B_{\pi(i)}$ is a connected set of nodes in $\graph$, implying that the distance between any two nodes is at most $\delta-1$. This in turn, implies that the distance between any two nodes in $C_i$ is at most $\delta-1$. 
  
  Therefore, the security state of a player $u$ can switch from being unsecured to secured due to the addition of another player $v$,  who is at most at a distance of $\delta$ from $u$. Conversely, any player $v$ can flip the security states of player $u$ which is at a distance of at most $\delta$ away from them. In summary, the marginal value added by the addition of a player $v$ is a function of the security costs of nodes that are at most $\delta$ away from $v$, which makes the allocation $(\delta+1)$-laterally implementable. 
   \hfill $\square$
\medskip

The $\delta$-agreeable allocation is the average of the set of all $\delta$-agreeable permutations. By considering the average across all permutations, we obtain symmetry, in exchange of polynomial-time computability.  

\noindent\textbf{Proof of \cref{thm_deltaagreeable}.}
  (i), (iii), (iv), and (v) follow from the fact that it is a convex combination of extreme core allocations satisfying these properties as demonstrated by \cref{prop:ec1}. The allocation is symmetric, because if there are two players $u$ and $v$ such that $c(S\cup \{u\}) = c(S\cup \{v\})$ for every $S \subseteq N\setminus \{u, v\}$, then from the definition of $\delta$-MRS, the following holds. \emph{If $S \cup \{u\}$ is a $\delta$-MRS for some $S \subseteq N\setminus \{u, v\}$, then so is $S \cup \{v\}$. } Thus, for every $\delta$-agreeable permutation of the form $(v_1, v_2, \dots, v_i, u, v_{i+1}, \dots, v_j,v, v_{j+1},\dots)$, $(v_1, v_2, \dots, v_i, v, v_{i+1}, \dots, v_j,u, v_{j+1},\dots)$ is also a $\delta$-agreeable permutation. Thus, averaging over the corresponding extreme core allocations implies equal payoffs for both $u$ and $v$.
\hfill $\square$
\medskip

\noindent\textbf{Proof of \cref{deltaagreeable_existence}.}

\noindent i. For any subset $S\subset N$, a $\delta$-MRS is also a $(\delta+1)$-MRS. Thus the existence of a $\delta$-agreeable allocation guarantees that of $(\delta+1)$-agreeable allocation. 
\smallskip

\noindent ii. Consider a complete graph $\graph=(V,A)$ of size $n$. For each $v\in V$, let $\theta_v = 0$. For each $u,v \in V$, let  $\xi_{uv} = 1$. Let $L_v = n-\delta-1$. 
    Now, any subset $S\subseteq N$ with $\delta$ or fewer players have no incentive to secure themselves. 
    Because, each player in $S$ will have to secure itself from the extrinsic risk from the players in $N\setminus S$. But if $|S|\le \delta$, $|N\setminus S| \ge n-\delta$. Given each $\xi_{uv} = 1$, each player incurs a cost of $n-\delta$ to secure itself, while the expected penalty from being unsecured is only $n-\delta-1$. Thus, there exist no $(\delta-1)$-MRS. Therefore, a $(\delta-1)$-agreeable allocation does not exist. On the other hand any set of $\delta+1$ players have an incentive to secure themselves in the above example, implying the existence of a $(\delta+1)$-agreeable allocation. 
    \smallskip
    
\noindent iii.  An $n$-agreeable allocation always exists because by definition, we consider the scenario where all players in the network-optimal security strategy are secured.  
\smallskip 

\noindent iv. Suppose $\delta$-agreeable allocation does not exist for $\delta=n-1$. 
    This necessarily implies that all $n$ players in the first iteration of \cref{alg:rhoAgree}. 
    This means all $n!$ permutations of the $n$ players are valid permutations. 
    Now, it is a well known result that the average of all $n!$ extreme allocations in cooperative game is the Shapley value. To prove the converse, we show that at least one of the $n!$ permutations is not included in the averaging if $\delta$-agreeable allocation exists for some $\delta \le n-1$.
    If $\delta$-agreeable allocation exists for some $\delta \leq n-1$, then, it necessarily means that all $n$ players were not added to the permutation $\varphi$ in the first iteration of \cref{alg:rhoAgree} since the first iteration can only add at most $\delta$ players.  This indicates that there were at least two iterations within \cref{alg:rhoAgree}. 
    Now, let $v_1$ be a node added in the first iteration of \cref{alg:rhoAgree} and $v_2$ be a node added in the second iteration. Any permutation starting $\varphi = (v_2, v_1, \dots)$ is not a valid permutation, and hence is not included in the averaging. Thus, the $\delta$-agreeable permutation cannot be the Shapley value. Further, from (i), it follows that the $n$-agreeable permutation coincides with the $\delta$-agreeable permutation and therefore, cannot be the Shapley value either. 
\hfill $\square$
\medskip

Now, we illustrate the computation of the $\delta$-agreeable allocation with an example, and also clarify the notion of $\delta + 1$-lateral implementability. 

\medskip 

\begin{example}[$\delta$-agreeable allocation]\label{example2} 

Consider the network $\mathbb{G}$, as depicted in \cref{figec2}, with the player set $N = \{ 1, 2, 3, 4, 5\}$ and the arc set $A = \{ (1,3), (2,3), (3,2), (3,4), (4,3), (3,5)\}$. Let $\theta_i = 10$ and $L_i = 20$ for all $i \in \{ 1, 2, 3, 4, 5\}$. Further, let $\xi_{13} = \xi_{35} = 5$, whereas $\xi_{23} = \xi_{43} = 7$ and $\xi_{32} = \xi_{34} = 14$.  

\begin{figure}[!htb]
  \begin{center}
    \includegraphics[scale = 0.45]{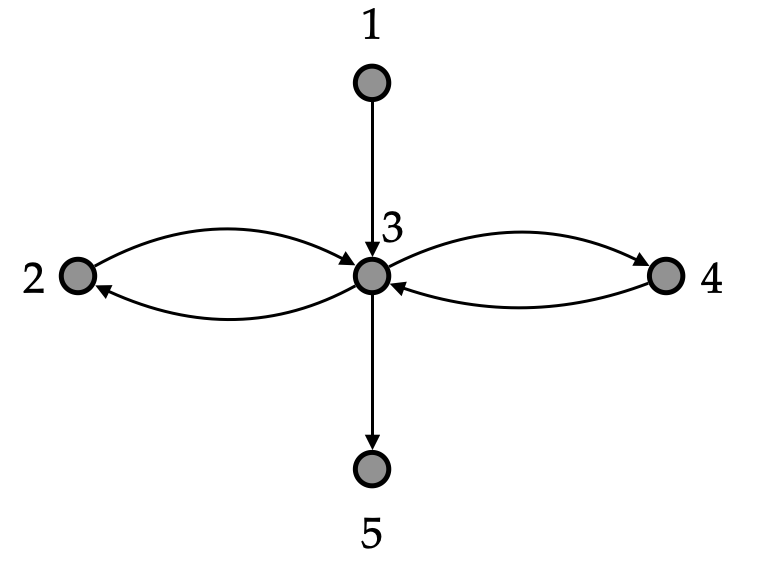}
    \caption{Multilateral implementability: an illustrative example}
    \label{figec2}
  \end{center}
\end{figure}

First, it is easily seen that in this network, the agreeable allocation does not exist. To see this, observe that for each of the players $\{1, 5\}$, it is individually rational to secure themselves and therefore, $\mathcal{S}_1 = \{1, 5\}$. However, $\mathcal{S}_2$ is empty because players $2$, $3$, and $4$ will not be secured even conditional on being in a coalition with $\{1, 5\}$. Therefore, $\mathcal{S}_2$ is empty implying there does not exist an integer $l$ such that $\bar{\mathcal{S}}_l = N$. Thus, the agreeable allocation does not exist. 

Now, let us consider the $\delta$-agreeable allocation, and we shall demonstrate that it exists for $\delta = 2$. First, we construct all $\delta$-agreeable permutations of the players in $N$ according to \cref{alg:rhoAgree}. To do so, note that for $S_0 = \emptyset$, the $\delta$-minimal rational security sets, $\mathcal{R}_{\delta}(S_0)$, are exactly the singleton sets, $\{1\}$, and $\{5\}$ since it is individually rational for these players to secure themselves. Thus, $S_1 = \{1, 5\}$. Then, the $\delta$-minimal rational security sets for the coalition $S_1$, $\mathcal{R}_{\delta}(S_1)$ is achieved for $\delta = 2$, and consists of the sets $\{2, 3\}$ and $\{3, 4\}$, since it is rational for players $2$ and $3$ (and  $3$ and $4$) to both be secured when they are jointly in a coalition with $\{1, 5\}$. Thus, the set of valid permutations of players $2$, $3$, and $4$, from \cref{alg:validPerm} are the ordered sets: $\{2,3,4\}$, $\{3,2,4\}$, $\{3,4,2\}$, and $\{4,3,2\}$.    

Therefore, following \cref{alg:rhoAgree}, the set of $\delta$-agreeable permutations of all players in $N$ are exactly the ordered sets: $\{1, 5, 2,3,4\}$, $\{1, 5, 3,2,4\}$, $\{1, 5, 3,4,2\}$, $\{1, 5, 4,3,2\}$, $\{5, 1, 2,3,4\}$, $\{5, 1, 3,2,4\}$, $\{5, 1, 3,4,2\}$, $\{5, 1, 4,3, 2\}$. 

By considering and averaging the extreme core allocations corresponding to each agreeable permutation, as depicted below, we obtain the $2$-agreeable allocation. 

\begin{figure}[!htb]
  \begin{center}
    \includegraphics[scale = 0.42]{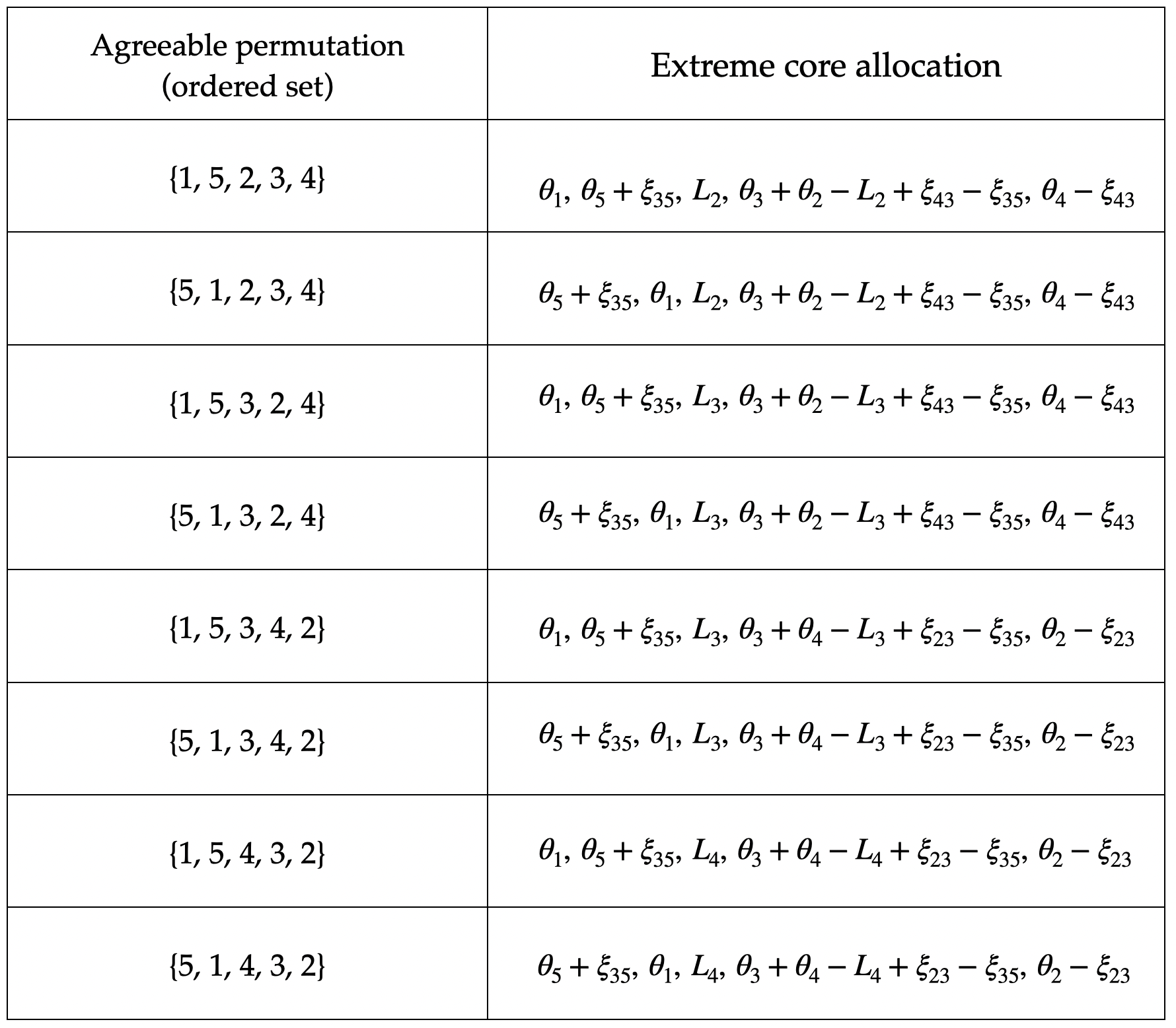}

    \label{tableec2}
  \end{center}
\end{figure}

Therefore, the $2$-agreeable allocation is given by, 
\begin{align*}
x^*_1 &= \theta_1 \\
x^*_2 &= L_2/4 + 3\theta_2/4 +\theta_3/4 - L_3/4 + \xi_{43}/4 - \xi_{35}/4 - \xi_{23}/2\\
x^*_3 &= L_3/2 + \theta_3/2 + \theta_2/4 + \theta_4/4 - L_2/4 - L_4/4 + \xi_{43}/4 + \xi_{23}/4 - \xi_{35}/2\\
x^*_4 &= L_4/4 + 3\theta_4/4 +\theta_3/4 - L_3/4 + \xi_{23}/4 - \xi_{35}/4 - \xi_{43}/2\\
x^*_5 &= \theta_5 + \xi_{35} 
\end{align*}

To observe that the $2$-agreeable allocation is not bilaterally implementable, notice that the cost allocated to player $2$, $x^*_2$ includes terms $\xi_{43}$ and $\xi_{35}$ that do not involve the player $2$. Similarly, $x^*_4$ includes terms $\xi_{23}$ and $\xi_{35}$ that do not involve player $4$. Thus, the cost allocated to players $2$ and $4$ is not expressible in the form of equation \ref{eqn7}. However, it is trilaterally implementable. 

\end{example}
\newpage

\section{Public Information Model and Quasi-Homogeneous Networks}
\medskip 

The following algorithm formalizes the equilibrium selection procedure for the scenario when all players in $N$
 are in independent coalitions. 
 
\begin{algorithm}[h]
\begin{algorithmic}
\Ensure{$\hat{\Upsilon}^i_{\{i\}; \rho}$ for $i \in N$}
\State $X \gets \emptyset$\;
\State $k \gets 1$\;
\While{$k \neq 0$}
\State$ Y \gets \emptyset $\;
  \For{$i \in N\backslash X$}
  \If{$L_i \geq \theta_i + \Sigma_{j \in N^-(i)\backslash X} \xi_{ji}$}
      \State $\hat{\Upsilon}^i_{\{i\}} = 1$, 
     \State$X \gets X \cup \{i\}$, 
     \State$Y \gets Y \cup \{i\}$, 
     \State$k \gets |Y|$.
     \EndIf
    \EndFor
\EndWhile
\end{algorithmic}
\caption{Independent security strategy under public information}\label{alg:public_independent}
\end{algorithm}

\begin{lemma}\label{lemma:public_independent}
Given the coalition structure $\rho$ with all players in independent coalitions, \cref{alg:public_independent} computes an equilibrium security state of player $i$, $\hat{\Upsilon}^i_{\{i\}; \rho}$, in polynomial time. 
\end{lemma}

\noindent\textbf{Proof of \cref{lemma:public_independent}.} First, we note that it is easy to see \cref{alg:public_independent} terminates in a polynomial number of steps. At some iteration, if the set $X$ does not change, then the algorithm terminates. Identifying the players in $N\setminus X$ to add to set the $X$ in each step involves checking a simple condition and since, the size of the set $N\setminus X$ strictly reduces in every step except the terminating one, the algorithm can proceed for at most $|N|$ steps. 

Consider a player $i \in N$ and let $\hat{\Upsilon}^i_{\{i\};\rho}$ denote the security state for player $i$ obtained upon termination of \cref{alg:public_independent}. We now show that $\hat{\Upsilon}^i_{\{i\};\rho}$ is an equilibrium security state for player $i$. Suppose $\hat{\Upsilon}^i_{\{i\};\rho} = 0$.  That is player $i$ is unsecured once the algorithm terminates. Note that upon termination of the algorithm, the set $X$ denotes the set of secured players. Then, suppose that player $i$'s security cost would be lowered by securing itself. Then, it must be that $L_i \geq \theta_i + \Sigma_{j \in N^-(i)\backslash X} \xi_{ji}$, but if this were so, then the algorithm would have assigned $\hat{\Upsilon}^i_{\{i\};\rho} = 1$, yielding a contradiction. Similarly, if suppose $\hat{\Upsilon}^i_{\{i\};\rho} = 1$.  That is player $i$ is secured once the algorithm terminates. Then, since the algorithm assigned $\hat{\Upsilon}^i_{\{i\};\rho} = 1$, $L_i \geq \theta_i + \Sigma_{j \in N^-(i)\backslash X} \xi_{ji}$. If player $i$ were instead unsecured, then the cost incurred by $i$ is $L_i$ which would not be lesser than the cost incurred by player $i$ under the current outcome. Therefore, \cref{alg:public_independent} terminates in polynomial time and computes an equilibrium security state of $i$ for all players $i$ in $N$.
\hfill $\square$
\medskip

We now extend the equilibrium selection procedure described in \cref{alg:public_independent} to compute an equilibrium security strategy for a coalition $S \subseteq N$ with a general partition $\rho$ of $N$ with $S \in \rho$.

\begin{algorithm}[h]
\begin{algorithmic}
\Ensure{$\hat{\Upsilon}^i_{S; \rho}$ for $i \in S \subseteq N$}
\State $X \gets \emptyset$\;
\State $k \gets 1$
\While{$k \neq 0$}
\State $ Y \gets \emptyset $
  \For{$S \in \rho$}
  \State  minimize $\Sigma_{i \in S} \left( L_i(1-\hat{\Upsilon}^i_{S; \rho})+ \theta_i\hat{\Upsilon}^i_{S; \rho} + \sum\limits_{(j, i) \in A,\ j \in (N \setminus X)} \xi_{ji}\right)$\;
 \State $Y \gets Y \cup \{i : \hat{\Upsilon}^i_{S; \rho} = 1\}$,
 \State  $k \gets |Y|$
 \State  $X \gets X \cup \{i : \hat{\Upsilon}^i_{S; \rho} = 1\}$,
\EndFor
\EndWhile
\end{algorithmic}
\caption{An equilibrium security strategy for a coalition under public information}\label{alg:public_coalition}
\end{algorithm}

\begin{lemma}\label{lemma:public_coalition}
Given a general coalition structure $\rho$, \cref{alg:public_coalition} computes an equilibrium security state of player $i$ in coalition $S$, $i$,$\hat{\Upsilon}^i_{S; \rho}$, in polynomial time.
\end{lemma}

\noindent\textbf{Proof of \cref{lemma:public_coalition}.} As in the proof of  \cref{lemma:public_coalition}, we note that it is easy to see \cref{alg:public_coalition} terminates in a polynomial number of steps since in each iteration either the size of the set $N\setminus X$ strictly reduces or the algorithm terminates. Further, each step in the algorithm involves minimizing $\Sigma_{i \in S} \left( L_i(1-\hat{\Upsilon}^i_{S; \rho})+ \theta_i\hat{\Upsilon}^i_{S; \rho} + \sum\limits_{(j, i) \in A,\ j \in (N \setminus X)} \xi_{ji}\right)$. The objective of the minimization problem can again be mapped on to the weight of a minimum directed cut separating the node set $\left(N\setminus S\right)\setminus X$ from the node set $l \cup (X\setminus S)$ in the auxiliary directed graph $\mathbb{G}^*$. Thus, the minimization problem can be solved also in polynomial time. Therefore, \cref{alg:public_coalition} runs in polynomial time. The proof that $\hat{\Upsilon}^i_{S; \rho}$ is an equilibrium outcome is identical to the arguments in the proof of  \cref{lemma:public_independent} and thus we omit them here. \hfill $\square$
\medskip

\noindent\textbf{Proof of \cref{public_core}.} We provide a proof by example. Let $N=\{1,2,3\},\,A=\{(1,2), (1,3)\}$. Let $\theta_1 = \theta_2 = \theta_3 = 10$, $L_1 = 0$, $L_2 = L_3  = 100$. Further, let $\xi_{12} = \xi_{21} = 20$. It is easily seen that the network-optimal security strategy secures all players and that players $2$ and $3$ will compensate $1$ for securing itself. However, player $3$ can defect from the grand coalition knowing that in the coalition structure, $\{\{1,2\}, \{3\}\}$, player $1$ will still be secured and be compensated by player $2$. Thus, the grand coalition is not stable to defections. \hfill $\square$
\medskip

The following \Cref{alg:publicAgreeable} computes the agreeable allocation $\hat{x}$. The algorithm takes in the family of sets $\mathcal{T}$ constructed in \S6 as an input. 

\begin{algorithm}[h]
\begin{algorithmic}
\Ensure{$\hat{x}_i$ for $i \in N$}
\For{$k \in \mathbb Z$, $1 \leq k \leq \ell$}
  \If{$k$ is odd}
    \For{$i \in \mathcal{T}_{k}$}
      \State $\hat{x}_i \gets \theta_i + \Sigma_{j \in N\backslash \bar{\mathcal{T}}_{k}} \xi_{ji}$
    \EndFor
  \EndIf
  \If{$k$ is even}
    \For{$i \in \mathcal{T}_{k}$}
      \State $\hat{x}_i \gets \theta_i + \Sigma_{j \in N\backslash \bar{\mathcal{T}}_{k+1}} \xi_{ji} - \Sigma_{j \in \bar{\mathcal{T}}_{k-1}} \xi_{ij} + \Sigma_{j \in \mathcal{T}_{k}} \xi_{ji}/2 - \Sigma_{j \in \mathcal{T}_{k}} \xi_{ij}/2$
    \EndFor
    \For{$i \in \bar{\mathcal{T}}_{k-1}$}
      \State $\hat{x}_i \gets \hat{x}_i - \Sigma_{j \in {\mathcal{T}}_{k+1}} \xi_{ji}$
    \EndFor
  \EndIf
\EndFor
\end{algorithmic}
\caption{Computing the agreeable allocation $\hat{x}$ under public information}\label{alg:publicAgreeable}
\end{algorithm}

\noindent\textbf{Proof of \cref{thm:public_main}.} 

i. Note that under the independent coalition structure, each player $i$ is either unsecured and therefore incurs a cost $L_i$, or player $i \in \mathcal{T}_1$. 
The agreeable allocation $\hat{x}$ allocates to all players $i$ a payoff smaller than $L_i$ so if player $i$ is unsecured, then it is immediately better off under the agreeable allocation. 
Suppose $i \in \mathcal{T}_1$. Then, in the independent coalition structure, player $i$ incurs a cost $x_i = \theta_i + \Sigma_{j \in N\backslash {\mathcal{T}}_{1}} \xi_{ji}$. This is identical to the update equation for $\hat{x}_i$ in the first iteration of the algorithm since ${\mathcal{T}}_{1} =  \bar{\mathcal{T}}_{1}$. Then, note that the only other update equation for $\hat{x}_i$ is when $k = 2$ and $i \in \bar{\mathcal{T}}_{1}$. In this update equation, the assigned value for $\hat{x}_i$ either remains the same or is reduced. Therefore, player $i$ cannot be worse off with the agreeable allocation. This shows that the agreeable allocation is individually rational. That is, all players will prefer to remain in the grand coalition over forming the independent coalition structure.

ii. From \cref{lemma:public_coalition} and \cref{coopgame_cost}, it follows that the family of sets $\mathcal{T}$ can be computed in polynomial time. Further, clearly, there are at most $|N|$ sets in the family of sets $\mathcal{T}$. That is, $l \leq n = |N|$. Therefore, the algorithm runs for at most $n$ iterations. Further, within each iteration, the computation of $\hat{x}_i$ is trivial. Thus, \Cref{alg:publicAgreeable} computes the agreeable allocation in polynomial time.   

iii. The expressions for $\hat{x}_i$ in all three cases in \Cref{alg:publicAgreeable} (i.e., $k$ is odd and $i \in \mathcal{T}_k$, $k$ is even and $i \in \mathcal{T}_k$, or $k$ is even and $i \in \mathcal{T}_{k-1}$) only contains cost parameters pertaining to player $i$ or its partners $j \in N$. Therefore, clearly, the agreeable allocation is bilaterally implementable. 

iv. If two players $i$ and $j$ are symmetric, then they will belong to the same set $\mathcal{T}_k$ for some $k$. Therefore, the allocation received by $i$ and $j$ will also be identical and therefore, $\hat{x}$ is a symmetric allocation. 

v. If a player $i$ is such that its marginal contribution to any coalition is $0$, then, recall our assumption that all players are secured in the grand coalition. Then, considering the coalition $N\setminus \{i\}$, it follows that for player $i$, $\theta_i = \xi_{ij} = \xi_{ji} = 0$. Therefore, all update expressions for $\hat{x}_i$ in the algorithm also evaluate to zero. Thus, the agreeable allocation satisfies the null player property. This concludes the proof. 
\hfill $\square$
\medskip

\noindent\textbf{Proof of \cref{agreeable_homogeneous}.} Suppose $\graph$ does not contain a $k$-core where $k =$ {\footnotesize $\left\lceil \frac{L-\theta}{\xi} \right\rceil$}. Then, there exists $i \in N$ such that $|N^-(i)| \leq k$. That is, $|N^-(i)| \leq$ {\footnotesize$\frac{L-\theta}{\xi}$}. Therefore, $L \geq \theta + \xi|N^-(i)|$. Therefore, from \cref{prop_independent}, $i \in \coreSubset_{1}$, and thus, $\coreSubset_{1}$ is not empty. 

Suppose that $\coreSubset_{k-1}$ is not empty for $k > 2$. If $\bar{\coreSubset_{k-1}} = N$, then, we are done, since, an agreeable permutation exists. If not, suppose the subgraph induced by players in $N\backslash \bar{\coreSubset_{k-1}}$  be denoted by $\mathbb{H}$. Then, there again exists a player $i$ in $\mathbb{H}$ such that the in-degree of $i$ in $\mathbb{H}$, $|\left(N\backslash\bar{\coreSubset_{k-1}}\right) \cup N^-(i)| \leq k$. Note, from the proof of \cref{prop:BilateralSymmAlloc}, that $i \in \coreSubset_k$ if $L_i \geq \theta_i + \Sigma_{j \in N\backslash\bar{\coreSubset_{k-1}} \cap N^-(i)}\ \xi_{ji} - \Sigma_{j \in \bar{\coreSubset_{k-1}}\cap N^+(i)}\ \xi_{ij}$. Now, in our quasi-homogeneous network, we have that,
\begin{equation*}
\theta + \left(\Sigma_{j \in N\backslash\bar{\coreSubset_{k-1}} \cap N^-(i)}\ \xi\right) - \left(\Sigma_{j \in \bar{\coreSubset_{k-1}}\cap N^+(i)}\ \xi\right) \leq \theta + |N\backslash\bar{\coreSubset_{k-1}} \cup N^-(i)|\xi \leq L. 
\end{equation*}
The last inequality follows since $|N\backslash\bar{\coreSubset_{k-1}} \cup N^-(i)| \leq k$. Therefore, $i \in \coreSubset_{k}$ and the iterative procedure can continue. This completes the proof of part (i).

Now, suppose, $\graph$ contains a $(k,l)${\it-core}, denoted by $\mathbb{H}$, where $k = \ell +$ {\footnotesize $\left\lceil \frac{L-\theta}{\xi} \right\rceil$}. Consider $i \in \mathbb{H}$. Suppose $i \in \coreSubset_{k}$ for some $k$. Therefore, $i \in N\backslash\bar{\coreSubset_{k-1}}$. Then, 
\begin{equation*}
\theta + \left(\Sigma_{j \in N\backslash\bar{\coreSubset_{k-1}} \cap N^-(i)}\ \xi\right) - \left(\Sigma_{j \in \bar{\coreSubset_{k-1}}\cap N^+(i)}\ \xi\right) > \theta + k\xi - l\xi = \theta + {\frac{L-\theta}{\xi}}\xi \geq L. 
\end{equation*}
Thus, $i \notin \coreSubset_{k}$. This yields a contradiction. That is, $i$ does not belong to $\coreSubset_{k}$ for any $k$ and therefore, an agreeable permutation (equivalently, the agreeable allocation) does not exist.
\hfill $\square$
\bigskip

\newpage

\section{Analysis of the Partial Information Model}
\medskip

We now consider the more general {\it partial information model} wherein for a subset of firms $\mathcal{P} \subseteq N$, the security cost parameters and actions for $i \in \mathcal{P}$ are publicly known to all other firms in $N$. Whereas, for firms not in $\mathcal{P}$, their costs and actions are only privately known to themselves. Therefore, in this scenario, the information set for a player $i$ acting independently is given by, $I(i, \{i\}) = \lbrace \theta_j, \xi_{jk}, \xi_{kj}, L_j, x_j, y_{jk} : j \in \mathcal{P}\cup\{i\}, k \in N^-(j)\rbrace$. This, as described in \S2, is a generalization of the private and public information models discussed in the main paper. When $\mathcal{P} = \emptyset$, then we recover the private information model, and when $\mathcal{P} = N$, we recover the public information model. 

First, we describe the independent and coalitional security strategies for firms in this partial information model. As in the public information setting, this again poses some challenges. Since the actions and costs of some players in $N$ are publicly known, the security actions of any player (or of a coalition) in the network, in general, depends on the security actions of other players (or other coalitions). Thus, we need to consider the coalition structure of players in the network in order to accordingly characterize the equilibrium security strategies of any given player or coalition. Second, as is often the case in network games with partial or full public information, there could be a preponderance of Nash equilibria. Thus, we need to also argue for the selection of a particular and justifiable equilibrium security strategy. We address both these issues in the subsequent discussion. 

To characterize the equilibrium security strategy of players when acting independently, we consider the coalition structure $\rho$ where all firms are in independent coalitions. Observe that, in $\rho$, firm $i$'s belief on the security state of other firms will be as follows: $i$ forms a worst-case belief on the security state of players $j \notin \mathcal{P}$ since $j$'s parameters and actions are privately known only to $j$, i.e., $\sigma_{ji} = 0$, whereas for $j \in \mathcal{P}$, $\sigma_{ji} = \sigma_j$. Let $\tilde{\Upsilon}^i_{\{i\}; \rho}$ be an indicator function denoting the equilibrium security state of player $i$ acting independently where $\rho$ is the coalition structure with all players in independent singleton coalitions. We present the following characterization of $\tilde{\Upsilon}^i_{\{i\}; \rho}$ which follows directly from \cref{lemma:public_independent} and \cref{prop_independent}. 

\begin{lemma}\label{lemma:partial_independent}
Consider the network $\mathbb{G}$ and the coalition structure $\rho$ with all players in independent coalitions. Then, 

\noindent i. Define $\mathbb{G}'$ as the induced subgraph of $\mathbb{G}$ on the node set $\mathcal{P}$. Further, in $\mathbb{G}'$, let $\theta'_j = \theta_j + \sum\limits_{i \in N\backslash \mathcal{P}}\xi_{ij}$ for $j \in \mathcal{P}$. Then, $\tilde{\Upsilon}^i_{\{i\}; \rho}$ for $i \in \mathcal{P}$ in $\mathbb{G}$ is computed by \cref{alg:public_independent} with the reduced network $\mathbb{G}'$ as the input. 

\noindent ii. Let $S$ denote the set of players in $\mathcal{P}$ for which $\tilde{\Upsilon}^i_{\{i\}; \rho} = 1$ according to (i.). Then, for $i \in N\backslash \mathcal{P}$, $\tilde{\Upsilon}^i_{\{i\}; \rho} = 1$ if and only if $L_i \geq \theta_i + \sum\limits_{j\in N\backslash \mathcal{P}} \xi_{ji} + \sum\limits_{j\in \mathcal{P}\backslash S} \xi_{ji}$. 

\end{lemma}
\medskip

\noindent\textbf{Proof of \cref{lemma:partial_independent}.} The central idea behind \cref{lemma:partial_independent} is as follows. \cref{lemma:partial_independent} operates in two steps. First, we consider the firms in $\mathcal{P}$. Since these firms are unaware of the costs and actions of other players in the network, these firms are operating in an environment which is identical to the network $\mathbb{G}'$ with full public information except that if a firm in $\mathcal{P}$ (which is the node set of $\mathbb{G}'$) chooses to be secured, then it must also bear the cost of securing itself from players not in $\mathcal{P}$ (in the original network), since for these firms $j \in N\backslash \mathcal{P}$, firm $i$ forms a worst-case belief, $\sigma_{ji} = 0$ that these firms are not secured. This is equivalent to the firm $j$ absorbing these costs into its cost of intrinsic security. Once, firms in $\mathcal{P}$ choose their actions, then, firms in $N\backslash \mathcal{P}$ can observe the actions of firms in $\mathcal{P}$ and accordingly solve for their equilibrium security states analogous to the independent security strategy in the private information model, except that now firms $i \in N\backslash \mathcal{P}$ need only secure themselves from unsecured firms in $\mathcal{P}$, i.e., the firms in $\mathcal{P}\setminus S$. This completes the proof.
\hfill $\square$
\medskip

For clarity, observe that when $\mathcal{P}$ is an empty set, the condition in \cref{lemma:partial_independent}(ii) coincides with the expression in \cref{prop_independent}. Therefore, not surprisingly, when  $\mathcal{P} = \emptyset$, then $\tilde{\Upsilon}^i_{\{i\}; \rho} = 1$ identifies exactly the set of players in $S_I$ as independently secured in the private information model. Further, similarly, if $\mathcal{P} = N$, then \cref{lemma:partial_independent}(i) coincides with \cref{lemma:public_independent}.

We now extend the ideas above to characterize the equilibrium security strategy and security states of players when acting in coalitions. That is, we consider a general coalition structure $\rho$ and a coalition $S \in \rho$ to obtain the equilibrium security states and actions of the players in coalition $S$.

\begin{algorithm}[h]
\begin{algorithmic}
\Ensure{$\tilde{\Upsilon}^i_{S; \rho}$ for $i \in S \subseteq N$}
\State $X \gets \emptyset$\;
\State $k \gets 1$
\While{$k \neq 0$}
\State $ Y \gets \emptyset $
  \For{$S \in \rho$}
  \State  minimize $\Sigma_{i \in S} \left( L_i(1-\tilde{\Upsilon}^i_{S; \rho})+ \theta_i\tilde{\Upsilon}^i_{S; \rho} + \sum\limits_{(j, i) \in A,\ j \in (N \setminus \mathcal{P})} \xi_{ji} + \sum\limits_{(j, i) \in A,\ j \in (\mathcal{P} \setminus X)} \xi_{ji}\right)$\;
 \State $Y \gets Y \cup \{i : \tilde{\Upsilon}^i_{S; \rho} = 1\}$,
 \State  $k \gets |Y|$
 \State  $X \gets X \cup \{i \in \mathcal{P} : \tilde{\Upsilon}^i_{S; \rho} = 1\}$,
\EndFor
\EndWhile
\end{algorithmic}
\caption{An equilibrium security strategy for a coalition under partial information}\label{alg:partial_coalition}
\end{algorithm}

\begin{lemma}\label{lemma:partial_coalition}
Given a general coalition structure $\rho$, under the partial information model, \cref{alg:partial_coalition} computes an equilibrium security state of player $i$ in coalition $S$, $i$,$\tilde{\Upsilon}^i_{S; \rho}$, in polynomial time.
\end{lemma}

\noindent\textbf{Proof of \cref{lemma:partial_coalition}.} As in the proof of  \cref{lemma:public_coalition}, we note that it is easy to see \cref{alg:partial_coalition} terminates in a polynomial number of steps since in each iteration either the size of the set $\mathcal{P}\setminus X$ strictly reduces. If the size of the set $\mathcal{P}\setminus X$ does not reduce in some iteration of the algorithm, then in the subsequent iteration, $k = 0$ because all firms for whom it was rational to be unsecured in the previous iteration will remain unsecured. Further, the objective of the minimization problem in each iteration can again be mapped on to the weight of a minimum directed cut separating two node sets in the auxiliary directed graph $\mathbb{G}^*$. Thus, the minimization problem can be solved also in polynomial time. Therefore, \cref{alg:partial_coalition} runs in polynomial time. Finally, since at each iteration, coalitional rationality is maintained by ensuring each coalition solves its cost minimization problem given the security states of all other players in the network, therefore, it follows that  $\hat{\Upsilon}^i_{S; \rho}$ will automatically be an equilibrium outcome when the algorithm terminates. 
\hfill $\square$
\medskip

We can then obtain the total security cost of a coalition $S$ belonging to a general coalition structure $\rho$ of $N$ in the partial information model, $\tilde{c}(S; \rho)$, as follows, 

\begin{equation}\label{EC11}
\tilde{c}(S; \rho) = \sum\limits_{i \in S} \left( L_i(1-\tilde{\Upsilon}^i_{S; \rho})+ \theta_i\tilde{\Upsilon}^i_{S; \rho} + \sum_{\substack{(j, i) \in A \\ \tilde{\Upsilon}^i_{S; \rho} = 1, j \in N\backslash \mathcal{P}}} \xi_{ji} + \sum_{\substack{(j, i) \in A, j \in \mathcal{P} \\ \tilde{\Upsilon}^i_{S; \rho} = 1, \tilde{\Upsilon}^j_{T; \rho} = 0}} \xi_{ji}\right),
\end{equation}

where $S$ and $T$ are (possibly identical) coalitions in $\rho$ with $i \in S$ and $j \in T$. That is, players in coalition $S$ who are secured pay the costs of securing the links with firms $j \in N\backslash \mathcal{P}$ since the security costs and actions of these firms are private information not known to $S$. Further, firms in $S$ that are secured also pay the costs of securing links to other firms $j \in \mathcal{P}$ that are not secured. 

For clarity, note that when $\mathcal{P} = \emptyset$, then, (\ref{EC11}) coincides with (\ref{coalitioncost}), and therefore, $\tilde{c}(S; \rho) = c(S)$. Likewise, when $\mathcal{P} = N$, note that  (\ref{EC11}) coincides with (\ref{eqn12}), and therefore, $\tilde{c}(S; \rho) = \hat{c}(S; \rho)$.
\medskip

Also, we note that  the example provided in the proof of \cref{public_core} is easily modified to also  demonstrate the instability of the grand coalition when $|\mathcal{P}| = 2$, with  only players $1$ and $2$ in $\mathcal{P}$, while player $3$'s parameters and actions are privately known. Then, the grand coalition will again not be stable and player $3$ will defect from the grand coalition. 

Further, as a corollary from the proof of \cref{lemma:partial_coalition}, we obtain the following. 

\begin{corollary}\label{ec4corr1}
    Consider the network $\mathbb{G}$ and $k$ interdependent security cost sharing games under partial information with $\emptyset \subseteq \mathcal{P}_1\subset \mathcal{P}_2 \ldots \subset \mathcal{P}_k \subseteq N$ where $\mathcal{P}_i$ denotes the set of players whose cost parameters and actions are known publicly in the $i^{th}$ game. Then, if the grand coalition is stable for some $i$ for $1 \le i \le k$, then the grand coalition is stable for all $j \leq i$. 

\end{corollary}

The contrapositive of the above statement confirms the basic insight that if the grand coalition is unstable at a certain level of public information in the network, the grand coalition will continue to remain unstable at higher levels of information provisioning in the network.  Again, as noted before, the instability of the grand coalition even with partially public information is, in general, driven by two factors: the reduced benefits of information acquisition from cooperative security, and the free-riding of firms on the security actions and cost-sharing of firms whose parameters and actions are known publicly.
\medskip

\subsection*{Agreeable Allocation in the Partial Information Model}
\medskip 

Naturally, this again motivates us to search for a cost-sharing mechanism that can support cooperative security. We show that once again we can extend the agreeable allocation to this general partial information setting while retaining several of its desirable properties. Notably, we prove that, analogous to \cref{thm:public_main}, the partial information version of the agreeable allocation, when it exists, satisfies individual rationality, a weaker notion of stability wherein each player is better off in the grand coalition (i.e., with full cooperation) as compared to the independent coalitions (i.e., no-cooperation) scenario.

As in the case of private and public information, for ease of exposition, we restrict our attention to networks where all firms are secured in the grand coalition. The algorithm to compute the agreeable allocation, in this case, is presented in \cref{alg7}, and once again involves as a first step the recursive computation of a finite family of mutually exclusive sets denoted here by $T$. Then, the agreeable allocation computed for a player depends on its membership in the family of sets.

\begin{algorithm}
\caption{Computing the agreeable allocation in the partial information model}\label{alg7}
\begin{algorithmic}

\State $\bar{T_0}\gets \emptyset$
\State $j\gets 1$
\While{true}
\State $T_{2j-1}
\gets {\textsc{Externality$\mathcal{P}$}}(\bar {T_{2j-2}})$
\State $\bar{T_{2j-1}} \gets \bar{T_{2j-2}} \cup 
T_{2j-1}$

\State $T_{2j} \gets 
\left \{
i \in N\setminus  (\mathcal {P} \cup \bar{T_{2j-1}}):  {\Upsilon}^i_{
\bar{T_{2j-1}} \cup \{i\}} = 1 
\right \}
$
\State $\tilde{x}_{i} \gets \theta_i + \Sigma_{k \in N\setminus (\bar T_{2j-1})} \xi_{ki} + \Sigma_{k \in T_{2j}} \xi_{ki}/2 - \Sigma_{k \in (\bar T_{2j-1})} \xi_{ik}$
\State $\bar{T_{2j}} \gets \bar{T_{2j-1}} \cup T_{2j}$
\If {$\bar {T_{2j}} = \bar {T_{2j-2}}$ }
\State \textbf{break}
\EndIf
\State $j \gets j+1$
\EndWhile
\If {$\bar{T_{2j}} = N$}
\State {Agreeable allocation $\tilde{x}$ exists and computed}
\Else 
\State {No agreeable allocation exists}
\EndIf
\medskip
 
\Function  {\textsc{Externality$\mathcal{P}$}} {$\bar {\mathcal {S}}$}
\State $\bar{T_0}\gets \emptyset$
\State $j\gets 1$
\While{true}
\State $T_{2j-1} 
\gets {\textsc{Independent{$\bar{\mathcal{P}}$}}}(\bar {T_{2j-2}})$
\State $\bar{T_{2j-1}} \gets \bar{T_{2j-2}} \cup 
T_{2j-1}$
\State $T_{2j} \gets 
\left \{
i \in \mathcal{P}\setminus (\bar {\mathcal {S}} \cup \bar{T_{2j-1}}):  {\Upsilon}^i_{\bar {\mathcal {S}} \cup \bar{T_{2j-1}} \cup \{i\}} = 1 
\right \}
$
\State $\tilde{x}_{i} \gets \theta_i + \Sigma_{k \in N\setminus (\bar S \cup \bar T_{2j-1})} \xi_{ki} + \Sigma_{k \in T_{2j}} \xi_{ki}/2 - \Sigma_{k \in (\bar S \cup \bar T_{2j-1})} \xi_{ik}$
\State $\bar{T_{2j}} \gets \bar{T_{2j-1}} \cup T_{2j}$
\If {$\bar {T_{2j}} = \bar {T_{2j-2}}$ }
\State \textbf{break}
\EndIf
\State $j \gets j+1$
\EndWhile
\State \Return $\bar {T_{2j-2}}$
\EndFunction
\medskip

\end{algorithmic}
\end{algorithm}

\begin{algorithm}
\begin{algorithmic}
\Function  {\textsc{Independent{$\bar{\mathcal{P}}$}}} {$\bar {\mathcal {S}}$}
\State $\bar{T_0}\gets \emptyset$
\State $j\gets 1$
\While{true}
\State $T_{2j-1} \gets {\textsc{Independent$\mathcal{P}$}}(\bar {T_{2j-2}})$
\State $\bar{T_{2j-1}} \gets \bar{T_{2j-2}} \cup T_{2j-1} $
\State $T_{2j} \gets 
\left \{
i \in N\setminus (\mathcal{P} \cup \bar {\mathcal {S}} \cup \bar{T_{2j-1}}): \hat {\Upsilon}^i_{\rho(\bar {\mathcal {S}} \cup \bar{T_{2j-1}}, \{i\})} = 1 
\right \}
$

\State $\tilde{x}_{i} \gets \theta_i + \Sigma_{k \in N\setminus (\bar S \cup \bar T_{2j})} \xi_{ki} + \Sigma_{k \in T_{2j}} \xi_{ki}/2 $

\State $\bar{T_{2j}} \gets \bar{T_{2j-1}} \cup T_{2j}$
\If {$\bar {T_{2j}} = \bar {T_{2j-2}}$ }
\State \textbf{break}
\EndIf
\State $j \gets j+1$
\EndWhile

\State \Return $\bar T_{2j-2}$
\EndFunction

\Function  {\textsc{Independent$\mathcal{P}$}} {$\bar {\mathcal {S}}$}
\State $\bar{T_0}\gets \emptyset$
\State $j\gets 1$
\While{true}
\State $T_j \gets \left \{ i \in \mathcal{P}\setminus (\bar {\mathcal {S}} \cup \bar{T_{j-1}}) : \hat \Upsilon^i_{\rho(\bar {\mathcal {S}} \cup \bar{T_{j-1}}, \{i\})} = 1 \right \}$
\State $\bar{T_j} \gets \bar{T_{j-1}} \cup T_j$
\If {$\bar {T_j} = \bar {T_{j-1}}$ }
\State \textbf{break}
\EndIf 
\State $j \gets j+1$
\EndWhile
\For{$i \in \bar {T_j}$}
\State $\tilde{x}_i \gets \theta_i + \Sigma_{j \in N\setminus (\bar {\mathcal {S}} \cup \bar{T_{j}})} \xi_{ji}$
\EndFor

\For{$j \in \bar {\mathcal {S}}$}
\State $\tilde{x}_j \gets \tilde{x}_j - \Sigma_{i \in \bar {T_j}} \xi_{ij}$
\EndFor

\State \Return $\bar {T_j}$
\EndFunction

\end{algorithmic}
\end{algorithm}

Note that, in the partial information case, since there is a set of players $\mathcal{P}$ for whom their costs and actions are public information, and the set of players in $N\backslash \mathcal{P}$ for whom their information is private, this implies there are separate routines to handle the players in each of these two sets. Further, within each of these two sets of players, we in turn have two distinct steps where in one step, players are identified for whom it is individually rational to secure themselves given the players already identified as secured, and in the other step, players are identified who will secure themselves for the direct positive externality they bestow on the players already secured. This is identical to equations (10) and (11) describing the computation of the agreeable allocation in the public information model. 

For brevity, in \cref{alg7}, we have combined the construction of the family of sets $T$ as well as the agreeable allocation to each player $i \in T$. It can be seen that if $\mathcal{P} = \emptyset$, then the output of \cref{alg7} coincides with the agreeable allocation in the private information setting. If  $\mathcal{P} = N$, then the output of \cref{alg7} coincides with the agreeable allocation computed by \cref{alg:public_coalition} in the public information model.  

As, in the private information and public information models, when the construction procedure of the family of sets $T$ terminates, if the union of the sets does not comprise all the players in $N$, then the agreeable allocation does not exist. In the two results below, we demonstrate that versions of \cref{thm:public_main} and \cref{agreeable_existence} extend to the partial information model. In fact, naturally, since the partial information model is a generalization of the private and public information models, \cref{thm:partial_main} generalizes \cref{thm:public_main}.

\begin{theorem}\label{thm:partial_main} The agreeable allocation under partial information, $\tilde{x}$, computed by \cref{alg7}, when it exists, is (i) individually rational, (ii) polynomial-time computable, and (iii) bilaterally implementable. Further, it also satisfies, (iv) symmetry, and the (v) null player property. 
\end{theorem}

\noindent\textbf{Proof sketch of \cref{thm:partial_main}.} The key steps in the proof of \cref{thm:partial_main} mimic the proof of \cref{thm:public_main}. The agreeable allocation in the partial information model, as well, by construction, is guaranteed to be individually rational since each player $i$ is allocated at most its payoff in the independent coalition structure. Similarly, at each iteration, the size of the set $\bar{T}_j$ either increases or if it does not the algorithm terminates and therefore, the agreeable allocation is computed in polynomial time. Also, all the update equations involving $\tilde{x}_i$ only consist of terms involving players $i$ and partners $j$, therefore, again, by construction the agreeable allocation is bilaterally implementable. Further, all symmetric players will belong to the same set $T_j$ and hence will receive an identical allocation, thus, the agreeable allocation $\tilde{x}$ is also symmetric. 
\hfill $\square$
\medskip

Further, from the construction of the agreeable allocation in the partial information model, it follows that we can again comment on the existence of the agreeable allocation. Specifically, we note that the informational assumption does not play a role in the existence or non-existence of the agreeable allocation. Thus, for example, the discussion in our numerical case study \S8, wherein we analyze the existence of the bilaterally implementable agreeable allocation for real-world alliance networks with simulated parameters, remains unchanged regardless of the information model assumed in the network.

\begin{corollary}
Consider the network $\mathbb{G}$ and $k$ interdependent security cost sharing games under partial information with $\emptyset \subseteq \mathcal{P}_1\subset \mathcal{P}_2 \ldots \subset \mathcal{P}_k \subseteq N$ where $\mathcal{P}_i$ denotes the set of players whose cost parameters and actions are known publicly in the $i^{th}$ game. Then, if the agreeable allocation exists in the $i^{th}$ game for $1 \le i \le k$, then the agreeable allocation exists for the $j^{th}$ game for all $j$ in $1 \le j \le k$. 
\end{corollary}

\end{document}